\documentclass[11pt,letterpaper,notoc]{JHEP}
\preprint{UW/PT 01/10, UPR-941-T
} 

\title{Exact Results for Supersymmetric Renormalization\\
and the Supersymmetric Flavor Problem}
\author{Ann E. Nelson\\ 
Department of Physics, Box 1560, University of Washington,\\
           Seattle, WA 98195-1560, USA}
\author{Matthew J. Strassler\\
Department of Physics and Astronomy, University of Pennsylvania, \\
209 S. 33rd St., Philadephia, PA 19104, USA}

\abstract{We explore the effects of a strongly-coupled, approximately
scale-invariant sector on the renormalization of soft supersymmetry
breaking terms. A useful formalism for deriving exact results for
renormalization of soft supersymmetry breaking terms is given in an
appendix, and used to generalize previously known results to include
the effects of nontrilinear superpotential terms.  We show that a
class of theories which explain the flavor hierarchy without flavor
symmetries can also solve the supersymmetric flavor problem by
producing nearly degenerate masses for the first two generations of
scalar superpartners within each charge sector.  Effects from
trilinear scalar terms are also suppressed, although their initial
values must be relatively small.  Our mechanism results in testable
predictions for the superpartner spectrum.}

\newcommand{\bel}[1]{\be\label{#1}}
\def\yia{\llambda_{i a_1 . . . a_n}}
\def\yja{\llambda_{j a_1 . . . a_n}}
\def\yias{\llambda^{*i a_1 . . . a_n}}
\def\yjas{\llambda^{*j a_1 . . . a_n}}
\def\al{\alpha}

\def\be{\begin{equation}}
\def\ee{\end{equation}}
\newcommand{\eref}[1]{(\ref{#1})}
\newcommand{\Eref}[1]{Eq.~(\ref{#1})}
\newcommand{\rem}[1]{}
\def\tr{{\rm tr}}
\def\half{{1\over 2}}
\def\OO{{\cal O}}
\def\NN{{\cal N}}

\def\none{$\NN=1$}

\def\susy{supersymmetry}

\def\rarr{\rightarrow}

\def\betau{\beta_{\alpha^{-1}}}
\def\ZZ{{\bf Z}}
\def\XX{{\bf X}}
\def\SS{{\bf S}}
\def\TT{{\bf T}}
\def\RR{{\bf R}}
\def\DD{{\tilde D}}
\def\LLam{\hat\Lambda\!\!\!\!\hat\Lambda\!\!\!\!\hat\Lambda}
\newcommand{\VV}[1]{{\bf H}_#1}
\newcommand{\II}{{\bf I}}
\def\yy{{\bf \hat y}}

\def\QQ{{\cal Q}}
\def\bQQ{\bar{\cal Q}}

\def\tim{\tilde m^2}
\newcommand{\byd}[1]{\left({\partial \over\partial {#1}}\right)}
\newcommand{\dbyd}[2]{{\partial #1\over\partial #2}}
\newcommand{\ddbyd}[2]{{\partial^2 #1\over\partial #2^2}}
\newcommand{\ddbydd}[3]{{\partial^2 #1\over\partial #2\ \partial #3}}
\def\ta{\theta}
\def\susy{supersymmetry}

\def\alt{{}^<\!\!\!\!{}_\sim \ }
\def\SM{{\bullet}}
\def\Gammaprime{\Gamma\!\!\!\Gamma'}
\def\GeV{{\rm GeV}} 
\begin{document} 
 
\section{Introduction} 
 
Recently, we have shown that nearly-conformal strongly coupled
supersymmetric theories can explain the striking features of the quark
and lepton masses and mixing parameters, without flavor symmetry
\cite{Nelson:2000sn}. We also pointed out that this mechanism for
flavor could in some cases simultaneously solve the supersymmetric
flavor problem.  In this note we consider in more detail the effects
of such theories on soft supersymmetry breaking terms. Exact
results for the renormalization of such terms show that under certain
conditions, these theories may also suppress flavor changing neutral
currents.  Our results for the running of soft SUSY-breaking terms
near supersymmetric fixed points have testable implications for the
supersymmetric flavor problem and the spectrum of superpartners.  In a
lengthy Appendix, we introduce a unifying formalism for the derivation
of the necessary results, and generalize previous results
\cite{Hisano:1997ua,Avdeev:1997vx,Jack:1997pa,Jack:1998eh,Jack:1998iy,Karch:1998qa,Karch:1998xt,Kobayashi:1998jq,Giudice:1998ni,Arkani-Hamed:1998kj,Arkani-Hamed:1998wc,Luty:1999qc,Kobayashi:2000wk}
to include the effects of nontrilinear  superpotential
couplings.

\section{Summary of  Scenario for Flavor Hierarchy, and effects on 
soft terms} 
 
We begin with a review of the class of models introduced in 
ref.~\cite{Nelson:2000sn} to explain the hierarchy of fermion masses 
and mixing parameters. 
 
Besides the standard model gauge group, $S$, (possibly embedded in a grand 
unified gauge group), we assume another (not necessarily simple) gauge 
group $G$, which we call the ``conformal sector''.  
The matter content includes standard model fields $X$ charged under 
$S$, and additional fields $Q$, charged under $G$. Some $Q$ fields 
must also carry standard model charges. 

 It is the strong 
dynamics of $G$ which will generate the flavor hierarchy. 
In ref.~\cite{Nelson:2000sn}, 
we argued that in order to produce a fermion mass hierarchy of order 
$10^5$, $G$ must be strongly coupled over a large energy range. This 
is possible if the strong couplings of the theory are in the vicinity 
of an approximate infrared conformal fixed point (CFP). 
 
In order to escape the strong dynamics of $G$ at low energies, the 
theory must also contain some relevant term, which is small at short 
distance, but which eventually drives the theory away from the CFP. 
This will sometimes require an additional gauge group $H$ with 
additional matter; we will call this the ``escape sector''. 
 
Our scenario contains several energy regimes.  There may or may not be 
a ``pre-conformal regime'' between the Planck/string scale and a scale 
$M_>$, in which the conformal sector has not yet reached the vicinity 
of a CFP.  In this regime the dynamics may be very complicated and we 
know little about it.  Between the scales $M_>$ and $M_<$ the theory 
is in a ``conformal regime''; all strong couplings are nearly scale 
invariant, any relevant couplings are very small, and the couplings of 
the standard model are perturbative.  There is then an ``escape'' from 
the conformal regime at the scale $M_<$.  Below $M_<$ we are in the 
``SSM'' regime; the low energy theory is simply a supersymmetric 
standard model with the special properties we will outline below. 
This regime survives down to the TeV scale. 
 
Renormalization generates the flavor hierarchy.  There are no 
flavor symmetries, or flavor hierarchies, at the Planck or string 
scale.  Our key assumptions are (a) that all standard model 
Yukawa couplings at the Planck scale are allowed and are of order 1, 
with no special relations among them, and (b) that all soft 
SUSY-breaking masses in the standard model sector are generated above 
or near the top of the conformal regime, and are roughly of the same 
order, again with no special relations among them. 
 
The essential results of this paper stem from the  features
of renormalization in both the conformal and SSM regimes.  In particular, 
the conformal regime suppresses most (though in general not all) 
soft SUSY-breaking parameters, while the SSM enhances some 
of them, sometimes in flavor-blind fashion.  These two facts will 
make it possible to wash out much of the flavor dependence inherent 
in the SUSY-breaking parameters of hidden-sector models. 
 
More specifically, the conformal regime does the following: 
 
\begin{enumerate} 
\item It sets a number of supersymmetric couplings to fixed, strong values, 
and drives their supersymmetry-breaking higher components toward zero. 
\item It generates large anomalous dimensions, automatically positive, 
for certain standard model matter fields to which it is coupled in 
the superpotential. 
\item Some superpotential couplings, including certain Yukawa
couplings of standard model fermions to the Higgs boson, run toward
zero as a power of the renormalization scale, along with the
corresponding A-terms.  In particular, the fermions of the first two
generations and part of the third have their Yukawa couplings and
trilinear scalar couplings driven small.
\item Various linear combinations of scalar masses-squared also run
towards zero as a power of the renormalization scale. In some models,
this guarantees small masses for the first two generations of scalar
superpartners at the scale $M_<$.  These masses will not quite reach
zero due to small flavor-dependent violations of conformal symmetry
from the standard model interactions, and so the first two generations
of sfermions have small non-degenerate masses at $M_<$.
\item Other standard model couplings and soft SUSY-breaking parameters
are relatively unaffected.  In particular, the gaugino masses, and
masses and A-terms of some or all of the third generation scalars, remain
substantial.
\end{enumerate} 
After the escape from the conformal regime, the theory enters  
the SSM regime,  
in which 
\begin{enumerate} 
\item The standard model Yukawa couplings and most of the soft parameters 
undergo modest multiplicative renormalizations; the structure of 
the Yukawa couplings and A-terms is not changed significantly. 
\item The soft masses for light-generation sleptons and squarks,
driven small in the conformal regime, receive large, positive, and
flavor-blind additive renormalizations to their masses-squared.  By
the TeV scale, these dominate the flavor-violating masses-squared
which were present at $M_<$, leaving these sleptons and squarks with
masses that are nearly degenerate in each charge sector.
\item The soft scalar masses of at least part of the third generation,
including at least the top and left-handed bottom squarks, and the
Higgs doublets, remain of order $M_{SUSY}$, as do the A-terms for
third generation scalars.  We have no predictions for these
quantities, and take them to be free parameters.
\end{enumerate} 
 
These two regimes combine together to suppress flavor changing neutral 
currents. The spectrum of the model is distinctive.  
The sleptons and squarks of the 
first two generations, and perhaps part of the third,  show degeneracy in 
each charge sector,  explaining the absence of large flavor 
changing neutral currents from virtual superpartner exchange. 
As we will see, there are definite predictions for certain 
combinations of these masses as a function of the gaugino masses and 
the scale $M_<$.   
 
The running in the conformal regime ensures that the A-terms of the 
light generations, potential additional sources of flavor violation, 
will also end up small --- roughly proportional to the corresponding 
Yukawa couplings.  However, such A-terms are 
problematic \cite{Masiero:1999ub} for $\mu\rightarrow e\gamma$ and 
for electric dipole moments.  If our scenario is correct, then there is 
some additional overall suppression of  trilinear couplings. 
It is likely we will see observable flavor and 
CP-violating effects in the relatively near future. 

\section{The Conformal Regime} 
 
Many of our claims about the conformal regime have already been given 
in \cite{Nelson:2000sn}.  We argued that among all possible superpotential 
terms coupling standard model fields to the conformal sector, only 
terms {\it linear} in the standard model fields could be nonvanishing at the fixed point, 
with all others being irrelevant.  (This is the reason that some 
conformal sector fields $Q$ must also be charged under the standard 
model for our scenario to operate.) Those standard model fields $X$ which 
couple in this way to the conformal sector develop positive anomalous 
dimensions of order one.  These anomalous scalings suppress all 
of the other couplings of the $X$ fields, 
such as their Yukawa couplings to the 
Higgs bosons.   
 
In a superconformal field theory, there is a $U(1)$ R-symmetry which
determines operator dimensions, including the dimensions of the
standard model fields.  In our scenario, the conformal symmetry only
occurs below the gravitational scales, so the R-symmetry is an
accidental symmetry determined by the near-conformal dynamics.  Within
each standard-model charge sector --- say, among the three
electroweak-singlet leptons --- distinct R-charges $r_i$ will be held
by three linearly independent combinations of the three fields with
that standard model charge.  These three combinations are determined
dynamically and cannot be simply read off from the Lagrangian at the
Planck scale.  Each combination has a distinct dimension $d_i=3r_i/2$
in the conformal regime and cannot mix with the others; consequently,
its Yukawa couplings will all be reduced by a distinct suppression
factor, namely $(M_</M_>)^{d_i-1}$.  Thus the algebraic hierarchy in
the R-charges and dimensions leads to an exponential hierarchy in the
suppression factors, and a generational structure in the Yukawa
couplings automatically emerges even though none was present above
$M_>$.
 
A superconformal fixed point also suppresses many soft SUSY-breaking terms.
If a superconformal fixed point is stable, then all nonzero coupling
constants approach fixed values in the infrared.  A necessary
condition for this to be true involves a certain property of the beta
functions for all nonzero couplings in the vicinity of the fixed
point; the matrix
\bel{betamatrix} 
{\partial \beta_i\over\partial y_j} 
\ee 
must be positive definite, where $\beta_i$ is the beta function for 
coupling $y_i$ and the $y_i$ are the nonzero couplings at the fixed 
point.  The positivity of this matrix assures that the couplings 
flow toward, not away from, their fixed values in the infrared.  
 
However, a superconformal fixed point must also have no SUSY-breaking 
parameters; the coupling constants, thought of as background 
superfields, may have nonzero lowest components, but cannot have 
nonzero higher components. For example, the coupling superfield 
${\bf y} = y + \ta^2 A$, $y$ a Yukawa coupling and $A$ the corresponding 
A-term, must have fixed point value $y=y_*$, $A=0$.  This 
implies that stability requires not only that $y$ flow to $y_*$ but 
also that $A$ flow to zero.  Fortunately, when the couplings, beta 
functions and anomalous dimensions are all written as superfields, it 
becomes clear that these two conditions are the same; the positivity 
of the matrix 
\eref{betamatrix} ensures {\it both} that $y$ flows in the infrared to its 
fixed value {\it and} that $A$ flows to zero.  Said another way, 
the positivity of the matrix \eref{betamatrix} should be interpreted 
as a {\it superfield} condition; when expanded in components, it shows 
that all SUSY-violating components of coupling constants flow to zero. 
This result is shown in detail in an 
extended appendix, generalizing results of 
\cite{Karch:1998qa,Karch:1998xt,Arkani-Hamed:1998wc,Luty:1999qc,Cheng:1998xg}. 
 
Still, this is not enough to prove that {\it all} SUSY-breaking parameters 
flow to zero at a superconformal fixed point.  Certain combinations 
of soft scalar masses-squared will be higher components of coupling 
superfields, but other linear combinations will generally not be.  At 
the fixed point these 
orthogonal linear combinations generally do not flow at all.  
In particular, as explained in  Appendix B, if there is a non-R $U(1)$ global 
symmetry in the theory, then the sum of the soft masses-squared weighted 
by their $U(1)$ charges will not run at the fixed point \cite{Arkani-Hamed:1998kj}. 
 
In order to solve the supersymmetric flavor problem via 
renormalization, we must therefore consider models of fermion masses, 
along the lines of \cite{Nelson:2000sn}, in which the $U(1)_R$
symmetry of the superconformal algebra is uniquely determined for the
first two generations.   In Appendix B we argue that in such theories, the 
conformal dynamics will drive the soft masses of the 
lighter-generation sfermions toward zero. 
 
The lighter-generation scalar masses do not quite reach zero, due to 
violations of conformal symmetry in the standard model sector.  As 
shown in Appendix C, each scalar mass-squared approaches a 
flavor-dependent value of order 
$$ 
\tilde m^{2} \sim {\alpha_{3}\over 4\pi} M_{3}^2 
$$ 
where $\alpha_3$ and $M_3$ are the QCD coupling constant and gluino 
mass, and all quantities are to be evaluated at the scale $M_<$.  For 
a gluino mass of order $1$ TeV, which means $M_3\sim $ 400 GeV at the  
scale $M_<$, this gives $\tim\sim 10^3$ GeV$^2 \ll M_{SUSY}^2$.

In short, a well-chosen conformal sector will generate a 
fermion mass hierarchy and reduce all the A-terms and soft scalar 
masses for the light generations (and part of the third) to small 
values.  This generic result is the output which will serve as  
input for the SSM regime --- assuming that we can escape into it from the 
conformal regime. 
 
\section{Escape from the conformal regime} 
 
We now briefly consider the escape from the conformal regime.  There 
are only weak constraints on the scales $M_>$ and $M_<$.  The 
ratio $M_>/M_<$ must be sufficiently large that the Yukawa couplings 
and soft masses are driven to small values.  However, the ratio 
$M_</M_W$ must be sufficiently large that the SSM sector can do its 
work; the gaugino masses must have room to drive the scalar masses 
back to acceptably high values.  These constraints are not difficult 
to satisfy; typically $M_>\sim 10^{14-19}$ GeV, $M_< \sim 10^{10-16}$ GeV 
can lead to reasonable models.

How can the theory depart the conformal regime at the scale $M_<$? 
The most natural possibility is that some nontrivial dynamics occurring 
at that scale leads the group $G$ to disappear, perhaps through 
confinement.  Another possibility is that some dynamics causes the 
standard model to decouple from the conformal sector at the scale 
$M_<$, with the conformal sector surviving to lower energies but 
coupling only irrelevantly to the standard model.  In any case, the 
standard model fields $X$ must decouple from that sector at a high 
scale $M_<$, and the sector itself (which contains some fields $Q$ 
which are chiral and charged under $S$) must vanish well above the 
weak scale.  Specific models of the escape were discussed in 
ref. \cite{Nelson:2000sn}; suffice it to say that once the conformal 
sector is specified, it is neither easy nor impossible to construct 
such models.

On the other hand, the escape from the conformal regime must not ruin
the flavor near-degeneracy that the conformal regime has worked so
hard to build. In appendix D we will argue that in models with a
rapid escape from a conformal regime to a weakly-coupled regime,
the soft masses change only by factors of order one, and the small
nonsupersymmetric mass terms for the first two generations of scalars
remain small.

\section{The flow in the supersymmetric standard model regime} 
 
After the escape, the low-energy theory consists merely of a
supersymmetric version of the standard model; we will assume it is the
minimal one.  (Part of the conformal sector may still be around, but
with only irrelevant couplings to the standard model sector.)  The
running of A-terms is controlled by the corresponding Yukawa couplings
and by the A-terms themselves; for light fermions these are both
small.  The running of the gaugino masses and of the heavy-generation
Yukawa couplings and A-terms, and of the masses of the corresponding
squarks and possibly sleptons, are standard; these quantities may
change by a factor of order one. We assume that the soft
supersymmetry-breaking third-generation scalar masses are of the same
order as the standard model gaugino masses, as necessary for a viable
and natural model.\footnote{Even though the MSSM gauge and top fields
are not participating in the superconformal dynamics, we can imagine
hypotheses under which this assumption could be unwarranted. For
instance if the MSSM gauge and top Yukawa couplings are both strong at
the fundamental scale and run down to perturbative values at lower
energies, and if the soft terms are also present at the fundamental
scale, then the soft terms might be renormalized by an order of
magnitude or so, in unpredictable directions.} However, at $M_<$, the
first two scalar generations are lighter by a factor of order
$\sqrt{\alpha\over{4\pi}}$, where $\alpha$ is a standard model gauge
coupling.

After the conformal field theory decouples from the standard model, at
the scale $M_<$, the soft parameters run according to the usual MSSM
RG equations.  We will consider them at leading loop order.  The
gaugino masses $M_r$, $r=3,2,1$ for $SU(3)$, $SU(2)$ and $U(1)$, run
in the usual way, with $M_r/\alpha_r$ a one-loop renormalization-group
invariant.  The soft masses for the first two generations of sfermions
are more interesting. 
Since for these generations the Yukawa couplings and
A terms are small, there are only two important contributions at one
loop, one from gaugino masses and one from the hypercharge D-term.
These renormalizations are independent of the soft scalar masses
themselves, and so are additive.  Specifically, the mass-squared
$\tim_s$ of the field $\phi_s$ has beta function
$$ 
\beta_{\tim_s} = \sum_r \left|{M_r\over\alpha_r}\right|^2  
\ddbyd{\gamma_s}{(1/\alpha_r)} 
+{1\over 2 \pi} \alpha_Y Y_s \tim_Y 
$$ 
where $\gamma_s$ is the anomalous dimension of the field 
$\phi_s$, $Y_s$ is its hypercharge, and 
$$ 
\tim_Y = \sum Y_j\tim_j 
$$ 
summed over all matter particles.  (Here $\alpha_Y = (3/5)\alpha_1$,
the latter being normalized to unify with the other couplings.)
At one loop  
$$ 
\beta_{\tim_Y} =  {1\over 2 \pi} (\sum Y_j^2)\alpha_1 \tim_Y \  
= -{b_0^{(Y)}\over 2 \pi} \alpha_Y \tim_Y={\beta_{\alpha_Y}\tim_Y\over\alpha_Y}\ . 
$$ 
Here $b_0^{(Y)}$ is the
coefficient of the weak hypercharge one-loop beta function: $b_0^{(Y)}=-\sum_jY_j^2$.
Note that  $\tim_Y/\alpha_Y$ is RG invariant.  
We also have 
$$ 
\gamma_s = \cdots - \sum_r {1\over \pi} \alpha_r C_{sr} + \cdots 
$$ 
where $C_{sr}$ is the quadratic Casimir (Tr$(t^At^B)=C_s\delta^{AB}$, 
$C_s=(N^2-1)/2N$ for $SU(N)$ and $ q_s^2$ for U(1)) of the field 
$\phi_s$ under the gauge group $G_r$, whence 
$$  
\ddbyd{\gamma_s}{(1/\alpha)} = -{2\over \pi} \alpha_r^3 C_{sr} \ . 
$$ 
But 
$$ 
\beta_{\ln \alpha} = - \beta_{\ln \alpha^{-1}} = -\alpha \beta_{1/\alpha} 
= -{\alpha b_0\over 2 \pi} \Rightarrow \alpha = 
-{2\pi\beta_{\ln\alpha}\over b_0} 
$$ 
and 
$$ 
\beta_{\alpha^2} = 2\alpha^2 \beta_{\ln \alpha}  
= -{\alpha^3 b_0\over  \pi} \Rightarrow \alpha^3 = 
-{\pi\beta_{\alpha^2}\over b_0} 
$$ 
so 
$$ 
\beta_{\tim_s} =   \sum_r \left|{M_r\over\alpha_r}\right|^2 
{2\beta_{\alpha_r^2}\over b_0} C_{sr} 
-{Y_s\over b_0^{(Y)} } \beta_{\tim_Y} 
$$ 
which implies  
$$ 
\tim_s(\mu)-\tim_s(\mu_0) =  \sum_r \left|{M_r\over\alpha_r}\right|^2 
{2 C_{sr}\over b_0^{(r)}} [\alpha_r(\mu)^2-\alpha_r(\mu_0)^2] 
-{Y_s\over  b_0^{(Y)}}{\tim_Y\over \alpha_1} [\alpha_1(\mu)-\alpha_1(\mu_0)] 
$$ 
(Recall that in models with $SU(5)$ or greater unification, as well as 
in other models of supersymmetry breaking with $M_r\propto \alpha_r$, 
the factor $M_r/\alpha_r$ will be the same for the three standard model 
gauge groups.) 

Thus the soft mass-squared parameters for the sfermions at the
TeV scale take the form
\bel{softendpt} 
{\tim_s} = \sum_r  \left|{M_r\over\alpha_r}\right|^2 
{2C_{sr}\over b_0^{(r)}} [\alpha_r^2-\alpha_{r<}^2] +  
{Y_s\over  b_0^{(Y)}} {\tim_Y\over\alpha_1} [\alpha_{1<}-\alpha_1]  ;
\ee
plus a flavor-dependent piece of order $\alpha_{3<}M_{3<}^2/4\pi$. 
In this expression,
quantities with a subscript $<$ are evaluated at $M_<$ while
all others are evaluated at the scale of the sfermion masses. 
 
We cannot measure $M_<$ very well.  Consider $\tim_Y=0$.  Since 
 $\tim_{s<}$ is very small, 
$$ 
\left|\dbyd{(\tim_s/M_3^2)}{\ln M_<} \right| 
= \left|{\beta_{\tim_s}\Big|_{\mu=M_<}\over |M_3|^2} \right| 
\approx 
{\sum_r 2 \alpha_{r<}M_{r<} C_{sr}\over \pi M_3^2} = 
{\sum_r 2 \alpha_{r<}^3 C_{sr}\over \pi \alpha_3^2}  
$$ 
which is of order $1/100$.  Therefore we cannot measure $\ln M_<$ from 
the soft masses to better than about 10.  On the other 
hand, this means that the theory predicts certain 
combinations of squark and slepton masses. 
 
For squarks the term involving $\alpha_3$ is dominant, 
so the ratio of squark and gluino masses is  
$$ 
{\tim_{\tilde q}\over M_3^2}\approx {8\over 9}\left[1 - 
{\alpha_{3<}^2\over \alpha_3^2}\right] \ . 
$$ 
Since this is not very sensitive to $\ln M_<$, and since $M_<$ cannot
be taken too low, it is a prediction 
of the model that $\tilde m_{\tilde q}/|M_{\tilde g}| \sim .8-.9$.
More detailed predictions will require 
going beyond one loop.  
We can also predict the splitting 
between the different types of quarks.
The splitting 
between up-type and down-type $SU(2)$-singlet squark mass parameters is 
is 
\bel{udsplit} 
\tim_{\bar u}-\tim_{\bar d} \approx  
{2\over 33}\left[\left|{M_1\over\alpha_1}\right|^2 [\alpha_{1<}^2-\alpha_1^2] +  
{3\over2}{\tim_Y\over\alpha_1}  \left( \alpha_{1<}-\alpha_1\right)\right] 
\ , 
\ee 
while 
\bel{LRqsplit} 
\tim_q -{1\over 2}(\tim_{\bar u}+\tim_{\bar d}) \approx  
{3\over 2}\left|{M_2\over\alpha_2}\right|^2 [\alpha_{2<}^2-\alpha_2^2] -  
{1\over 22}\left|{M_1\over\alpha_1}\right|^2 [\alpha_{1<}^2-\alpha_1^2] 
 -  
{1\over 33}{\tim_Y\over\alpha_1}  \left( \alpha_{1<}-\alpha_1\right) \ . 
\ee 
Note that both of these splittings can be larger than the expected
small  splittings at the scale $M_<$.

For sleptons, consider the sum 
$ 
\tim_\nu + \tim_{e^-} + \tim_{e^+}. 
$ 
Since the hypercharges of these particles cancel, 
$$ 
{\tim_\nu + \tim_{e^-} + \tim_{e^+} \over 3 } =  
\left|{M_2\over\alpha_2}\right|^2 [\alpha_{2<}^2-\alpha_2^2]
+ 
{1\over 11}  \left|{M_1\over\alpha_1}\right|^2 [\alpha_{1<}^2-\alpha_1^2]
$$ 
Meanwhile,  
\bel{LRlsplit} 
\half(\tim_\nu + \tim_{e^-})- \tim_{e^+} 
\approx  
{3\over 2}\left|{M_2\over\alpha_2}\right|^2 [\alpha_{2<}^2-\alpha_2^2] -  
{3\over 22}\left|{M_1\over\alpha_1}\right|^2 [\alpha_{1<}^2-\alpha_1^2] 
 +  
{3\over 22}{\tim_Y\over\alpha_1}  \left( \alpha_{1<}-\alpha_1\right) \ . 
\ee 
This splitting can easily be much larger than that inherited from the scale $M_<$. 
It thus serves to measure $\tim_Y$, leading to predictions for 
the splittings in the squark sector, Eqs.~\eref{udsplit} and \eref{LRqsplit}.
 
Note that these formulas refer to soft supersymmetry-breaking mass
terms. The physical scalar masses will receive tiny corrections from
mixing and more important  corrections from electroweak symmetry breaking
via the quartic electroweak D-terms.

Unless R parity is violated, the splittings must be such that a
neutral particle, either the lightest neutralino or the sneutrino, is
the LSP.  Our scenario does not constrain the $\mu$ term, so it is
possible that the LSP is partly or altogether Higgsino; we will assume
$\mu \gg M_1$ for the following estimate.  Requiring that the
right-handed selectron and smuon be heavier than the Bino implies 
\be
\label{cosmo}
1<{\tim_{e^+}\over |M_1|^2} = {1\over 11}\left[2\left({\alpha_{1<}^2\over
\alpha_1^2} -1\right) -{\tim_Y\over |M_1|^2}\left({\alpha_{1<}\over
\alpha_{1}}-1\right)\right] -0.23{m_Z^2 \over |M_1|^2}\cos(2\beta)\ .
\ee 
(Here we have written the physical mass of the sleptons, not merely their
mass-squared parameter; the last term is due to the electroweak D-terms.)  For $M_< =
10^{16}$ GeV, \Eref{cosmo} requires that either $M_1 < 120 $ GeV or
$\tim_Y$ is negative. For $M_1 \gg M_Z$ we must have
$|\tim_Y|>1.3|M_1^2|$ for $M_<\sim M_{GUT}$. Lower $M_<$ will require
more negative $\tim_Y$.  Large negative $\tim_Y$ could make the
splitting in \Eref{LRlsplit} small, or negative, so the sneutrino
could be the lightest scalar and even the LSP.

\subsection{Examples}

Here we give a couple of sample spectra, in order to demonstrate
that there is a range  of viable possibilities
in which the $\tilde e^+$ is not
the LSP. Because our scenario
places no constraints on masses of third generation particles, we only
present results for the first two generations. We use one-loop
renormalization group equations, and neglect threshold corrections. 
All masses are renormalized at a  scale of 500 GeV.

For our first example, we take
$$
M_<=M_{GUT}=1.6\cdot 10^{16} \GeV \ , \quad
\alpha_{GUT} =1/24.5\ , \quad
M_{1/2}(M_{GUT}) = 400\ \GeV \ , 
$$
$$
{\rm D-term:} \ \ m^2_Y(M_{GUT})=-(300\ \GeV)^2\ .
$$
The Bino, Wino and gluino mass parameters at 500 GeV are
$$
171, \  332,\  1015\ \GeV
$$
and the approximate  masses for 
$\tilde u, \tilde d, \tilde {\bar u}, \tilde {\bar d}, 
\tilde \nu, \tilde e^-, \tilde e^+$ are
$$
\left((922\ \GeV)^2+0.35M_Z^2\cos(2\beta)\right)^{1\over2}, \
\left((922\ \GeV)^2-0.42M_Z^2\cos(2\beta)\right)^{1\over2}, $$ $$ 
\left((884\ \GeV)^2+0.15M_Z^2\cos(2\beta)\right)^{1\over2}, \
\left((882\ \GeV)^2 -0.07M_Z^2\cos(2\beta)\right)^{1\over2}, $$ $$ 
\left((278\ \GeV)^2+0.5M_Z^2\cos(2\beta)\right)^{1\over2},\  
\left((278\ \GeV)^2-0.27m_Z^2\cos(2\beta)\right)^{1\over2},\ $$ $$
\left((168\ \GeV)^2-0.23 M_Z^2\cos(2\beta)\right)^{1\over2}\  .
$$
For $\tan\beta >2$,  the $ \tilde e^+$ is not the LSP.
Small,  flavor non-degenerate, corrections to these masses come from the
nonvanishing of the masses at the scale $M_<$.  The natural size is 
$$
\Delta\tim\sim{\alpha_3M_3^2\over 4\pi}\Big|_{\mu=M_<}
\sim 520\ \GeV^2 \approx (23\ \GeV)^2\ ,
$$   
 a 0.06\% effect for the squarks and a 2\% effect for the sleptons.  (Note
the slepton splitting is dominated by $\alpha_3$, since the
order-one nonperturbative anomalous dimensions of lepton
superfields are generated by fields in the
conformal sector, which couple to $SU(3)$-color.  These do not
vanish in the limit $\alpha_1\to 0$, $\alpha_2\to 0$.  See Appendix C.)

If now we take a lower $M_<$, and continue to assume gauge coupling
and gaugino mass unification at the usual GUT scale\footnote{Above the
scale $M_<$, we cannot compute in a model independent way the effect of
the strong dynamics on the individual $\beta$ functions; however,
provided that the strong dynamics respects an approximate global SU(5)
symmetry, the gauge couplings and gaugino masses will still unify at
$1.6\times 10^{16}$ GeV. }, {\it e.g.}
$$
M_<=1.6\cdot 10^{12} \GeV \ , \quad
(\alpha_1,\alpha_2,\alpha_{3})(M_<) =(1/34.2,\ 1/26.0,\ 1/20.1)\ , 
$$ 
$$
(M_1,M_2,M_{3})(M_<)  = (215,283,366 )\ \GeV \ , \quad
$$
$$
{\rm D-term} \ \ m^2_Y(M_{<})=-(600\  \GeV)^2
$$
(note the large  $m^2_Y(M_{<})$, needed to avoid an
$\tilde e^+$ LSP) then
the Bino, Wino and gluino mass parameters  at 500 GeV are 
$$
129, \  250, \  762\  \GeV \ .
$$
Up to the aforementioned small nondegenerate corrections to the masses
from the scale $M_<$,  the masses for  
$\tilde u, \tilde d, \tilde {\bar u},
\tilde {\bar d}, \tilde \nu, \tilde e^-, \tilde e^+$ 
are
$$
\left((653\ \GeV)^2+0.35M_Z^2\cos(2\beta)\right)^{1\over2}, \
\left((653\ \GeV)^2-0.42M_Z^2\cos(2\beta)\right)^{1\over2}, \ $$ $$
\left((625\ \GeV)^2+0.15M_Z^2\cos(2\beta)\right)^{1\over2}, \
\left((634\ \GeV)^2 -0.07M_Z^2\cos(2\beta)\right)^{1\over2}, \ $$ $$
\left((146\ \GeV)^2+0.5M_Z^2\cos(2\beta)\right)^{1\over2}, \  
\left((146\ \GeV)^2-0.27m_Z^2\cos(2\beta)\right)^{1\over2}, \ $$ $$
\left((136\ \GeV)^2-0.23 M_Z^2\cos(2\beta)\right)^{1\over2} \ .
$$
Note the near degeneracy between the left- and right-handed
sleptons---in this case, for $\tan\beta > 1.7$,  the sneutrino is the
lightest scalar, although the Bino remains the LSP. 

As discussed in appendix C, the expected size of the  flavor
nondegeneracy in the mass-squared of all the scalars   is 
again of order 
$$
\Delta\tim\sim{\alpha_3M_3^2\over 4\pi}\Big|_{\mu=M_<}
\approx (23\ \GeV)^2\ ,
$$   
now an  2.6\% effect for the sleptons and an 0.1\% effect for the squarks.

 \section{Remnant Flavor Changing Neutral Currents} 
 
In the last section we computed the flavor-diagonal additive 
contribution that the light-generation scalar  soft masses obtain through 
standard model loops and the gaugino masses.  
More precisely, the soft masses should be written as matrices 
$\tim_{ij}$ in generation space.  In the basis where the
MSSM fields have definite anomalous dimensions at the infrared fixed
point,
all matrix elements except $\tim_{33}$ will be
small at $M_<$.  However, we must estimate more carefully their size.

We will imagine that supersymmetry breaking occurs through some
mechanism which generates gaugino masses of order $m_{1/2}$,
scalar masses of order $m_0$, and A terms of order $A_0$.
Either or both $m_{1/2}$ and $A_0$ could be naturally small, but we 
would not normally expect $m_0$ to be small.  Let us assume initially,
however, that all are of roughly the same order.

Assuming the  strong dynamics is consistent with SU(5)
symmetry,  at $M_<$
 the $\tim_{33}$
element of the $\tilde q,\tilde{\bar u},\tilde e$
matrices for will be of order $m_{0}^2$.  Unless $\tan\beta$ is
very large, at   $M_<$ all  elements of $\tilde\ell$ and 
$\tilde{\bar d}$,  including $\tim_{33}$, are small.

As previously noted and shown in  appendix C, at $M_<$
diagonal  mass matrix elements  for scalar partners of light fermions have 
unknown values (positive or negative) of order 
$\eta\equiv\alpha_{3<} M_{3<}^2/(4 \pi)$.  
In  appendix C we also show that off-diagonal factors are of order 
$$
\epsilon_{ij}=\OO\left( (M_</M_>)^{(d_i+d_j-2)}m_{0}^2\right)\ ,
$$ 
where $d_i$ is the
scaling dimension of the appropriate field in the conformal regime. 
Therefore, after running from $M_<$ 
to low energy,  the  scalar mass-squared matrices
 take the form  
\def\odeta{{\OO\left(\eta m^2_{1/2}\right)}} 
\bel{cfpmass}
\tim_{ij}\sim \left[\matrix{ 
\tim_s + \odeta & \epsilon_{12} & \epsilon_{13} \cr 
\epsilon_{12} & \tim_s + \odeta & \epsilon_{23} \cr 
\epsilon_{13} & \epsilon_{23} & \tim_{33}  } 
\right] \ . 
\ee 
Here $\tim_s$ is the appropriate flavor-independent
contribution for this charge of sfermion, as given in \Eref{softendpt}.
The element $\tim_{33}$ is  of order $m_{0}^2$ for type I fields and is
$\tim_s+\odeta$ for type II fields, where   type I fields  are those
whose third generation is weakly coupled to the
conformal sector, and type II are all others. Type I includes at least
$\tilde  t,\tilde b,\tilde{\bar t}$,  probably 
$\tilde\tau^+$,  perhaps $\tilde {\bar b}$, $\tilde \tau^-$ and
${\tilde \nu}_\tau$. 

The matrix in \Eref{cfpmass} is given in  the basis selected dynamically
by the CFP.    To study the resulting
flavor-changing physics, it is much more
convenient to work in the quark mass eigenstate basis. In
this basis the off-diagonal elements are dominated by  the residual
$\odeta$ effects. For type I fields, we have
\bel{massI}
\tim_{ij}(I)\sim \left[\matrix{ 
\tim_s + \odeta & V_{12}\odeta  &V_{13}\OO(m^2_{0}) \cr 
V_{12}\odeta & \tim_s + \odeta & V_{23}\OO(m^2_{0}) \cr 
V_{13}\OO( m^2_{0}) & V_{23}\OO(m^2_{0}) & \OO(m^2_{0})  } 
\right] \ , 
\ee 
 where $V$ is the matrix which rotates the CFP basis to the quark mass
basis; its off-diagonal entries are of order 
\bel{v}V_{ij}= \OO\left({M_<\over M_>} \right)^{|d_i-d_j|}\ .\ee
For type II fields, the off-diagonal entries involving the third
generation are smaller, and in the quark mass eigenstate basis we have
\bel{massII}
\tim_{ij}(II)\sim \left[\matrix{ 
\tim_s + \odeta & V_{12}\odeta  &V_{13}\odeta \cr 
V_{12}\odeta & \tim_s + \odeta & V_{23}\odeta \cr 
V_{13}\odeta & V_{23}\odeta & \tim_s + \odeta  } 
\right] \ , 
\ee 

The off-diagonal terms in these matrices will induce flavor-changing 
neutral currents (FCNCs) and violations of CP, and we must check that 
the models under discussion  are 
consistent with experimental constraints. 
 
We also have FCNCs and CP violation from the A-terms.  These inherit 
roughly the same structure as the Yukawa couplings.  
If unstructured at the scale $M_>$, at low energy they are of the form
\bel{a}
A_{ij}\sim \left({\rm Max}(y_i,y_j)\right)\left( {\rm
Max}(V^L_{ij},V^R_{ij}) \right)A_0\ , \ee
 where 
and $V_{ij}^{L,R}$ refer to
the left- and right-handed fields respectively. For the quark fields,
$V^L\sim V^{CKM}$. We expect $A_0$  to 
be of order $m_{0}$ or smaller, depending in detail on the 
supersymmetry breaking and communication mechanism. 
 
In determining the constraints on our scenario, we use the review of 
Gabbiani et al. \cite{Gabbiani:1996hi}, supplemented as necessary to 
incorporate  QCD corrections
\cite{Bagger:1997gg,Ciuchini:1998ix,Ciuchini:1998bw,Krauss:1998mt,Contino:1998nw,Buras:2000if,Feng:2000kg}.   We will 
see that  this
mechanism for flavor changing neutral current suppression is
roughly consistent with the constraints, although adequate suppression
of $\mu\rightarrow e \gamma$ and the electron electric dipole moment
constrains $A_0$ to be significantly smaller than $m_{0}$. 
 
\subsection{$K-\bar K$ system} 
\def\GeV{{\rm\ GeV}} 
\def\tc{\theta_C} 
  
 The $K_L$---$K_S$  mass difference and $\epsilon_K$ place constraints on the down-type
squark masses. For our scenario, the strongest constraints are
\begin{eqnarray}
\Big|{\rm Re}(\delta^d_{12})_{LL}(\delta^d_{12})_{RR}\Big|^{1/2} 
&< &1.2 \times 10^{-3}\left({m_{\tilde d}\over 500\GeV}\right) \cr 
\Big|{\rm Im}(\delta^d_{12})_{LL}(\delta^d_{12})_{RR}\Big|^{1/2} 
&< &0.1 \times 10^{-3}\left({m_{\tilde d}\over 500\GeV}\right) \ . 
\end{eqnarray} 
where,$(\delta^d_{12})_{LL(RR)}$ is the ratio of the off-diagonal term 
to the diagonal terms  in the
left(right)-handed squark mass-squared matrix, in the basis (see above) which
diagonalizes the down-type quark masses. We assume
the left- and right-handed $\tim_{12}$ are roughly of the
same order of magnitude---otherwise the constraints are much weaker.
The expected non-degeneracy in the squark masses is of order
$\alpha_s/(4 \pi)\sim 10^{-3}$, and the off-diagonal terms should
additionally be suppressed by mixing angles---here we assume both left- and
right-handed mixing angles to be of order the Cabbibo angle. Then both 
$(\delta^d_{12})_{LL}$ and $(\delta^d_{12})_{RR}$ are expected to be of order
$0.2 \times 10^{-3}$, potentially saturating  the
constraints coming from CP violation. 
 
Indirect evidence for a sizable supersymmetric contribution to
$\epsilon_K$
 might arise from
a B-factory measurement of the CP asymmetry in 
$B_d(\bar B_d)\rightarrow \psi K_S$. Should this asymmetry disagree with the standard
model prediction, one explanation would be that the CKM phase cannot
be deduced from $\epsilon_K$ using the standard model alone.

Much less constraining are the flavor changing  terms mixing the left-
and right-handed squarks, since these, which come from the
A-terms, are always suppressed by a factor of
order the strange quark Yukawa coupling times the Cabbibo angle.
$$ 
\Big|(\delta^d_{12})_{LR}(\delta^d_{12})_{RL}\Big|^{1/2}
\sim {y_s\over\sqrt{4\pi}}\tc  
<  2.4 \times 10^{-3}\left({m_{\tilde d}\over 500\GeV}\right) \ .
$$

\subsection{$B_d-\bar B_d$ Mixing}
Again,  for  $B_d-\bar B_d$ mixing, the strongest constraints are 
on the product of the off-diagonal terms in the product of the left- and right-handed squark
mixing matrices.
\bel{bmix}
\Big|(\delta^d_{13})_{LL}(\delta^d_{13})_{RR}\Big|^{1/2} 
< 2 \times 10^{-2}\left({m_{\tilde d}\over 500\GeV}\right) \ .
\ee 
While the nondegeneracy in this system may be of order 1, the  $\tim_{13}$
off-diagonal terms in the left-handed squark mass matrices will be suppressed by a
factor of order $V_{td}\sim 10^{-2}$.
In the right-handed matrices
there are two possible cases. The $\bar b$ could be a type I field, in
which case we expect
$$(\delta^d_{13})_{RR}\sim V_{13}^R\sim {m_d/ m_b\over V_{td}}$$
and 
$$ \sqrt{|(\delta^d_{13})_{LL}(\delta^d_{13})_{RR}|}(I)\sim \sqrt{m_d\over
m_b}\sim 3\times 10^{-2}$$
If the $\bar b$ is a type II field, then  $V^R_{13}$ is also suppressed by
a factor of $\eta\sim  10^{-3}$.

In both cases constraint  \eref{bmix} is easily
satisfied. For  type I right handed $b$ quark, and
 for 600 GeV or lighter squarks, there is a possible  $\OO(1)$ 
contribution  from supersymmetry to the phase of $B_d-\bar B_d$ mixing, which could show
up as an non-standard value for the time dependent  CP asymmetries in 
$B_d(\bar B_d)\rightarrow f_{CP}$ decays.

\subsection{$B_s-\bar B_s$ Mixing}

In the standard model the ratio of the $B_s$ mixing  to
 $B_d$ mixing is predicted to give approximately
$${\Delta m_{B_s}\over \Delta m_{B_d}}\Bigg|_{\rm
standard}\approx\Bigg|{V^{CKM}_{ts}\over V^{CKM}_{td}}\Bigg|^2\ \sim
20 \ .$$ 
This prediction should be tested at
upcoming measurements of the $B_s$ oscillation rate at the
Tevatron. It is therefore interesting to compare the ratio of the
flavor-violating supersymmetric contributions to the mixing, which is of order
$$
{(\delta_{23}^d)_{LL}(\delta_{23}^d)_{RR}\over
{(\delta_{13}^d)_{LL}(\delta_{13}^d)_{RR}}}
\sim 
{V^{d}_{23} V^{\bar d}_{23}\over V^{d}_{13} V^{\bar d}_{13}}
\sim
{m_s\over m_d}\sim 20  \ .$$ 
We therefore do not expect a big
correction to the standard model prediction for the ratio.
 The phase in the supersymmetric
contribution to $B_s$---$\bar B_s$ mixing could well be nonstandard,
however, and lead to an $\OO(1)$ CP-violating asymmetry in, {\it e.g.}
time dependent
 $B_s(\bar B_s)\rightarrow (J/\psi) \eta$, which would be a clear signal of
beyond the standard model physics.

\subsection{Constraints on A-terms from $\epsilon'/\epsilon, 
\mu\rightarrow e \gamma,$ and
Electric Dipole Moments}

 Our model falls into the class of models with A-terms of the same
texture as the Yukawa couplings, which were recently analyzed by
Masiero and Murayama \cite{Masiero:1999ub}. 
They found that for  500 GeV sleptons, 
  CP-violating A terms lead to an electron electric
dipole moment (EDM) of size comparable to the current bounds, and lepton flavor-violating A terms lead to  
  a $\mu\rightarrow e \gamma$ rate of order the experimental
bounds.  Note that with gaugino mass unification our scenario leads to 
sleptons which are much lighter than the gluino and squarks. Since natural electroweak
symmetry breaking places stringent upper bounds on the  gluino mass
\cite{Barbieri:1988fn,Dimopoulos:1995mi,Anderson:1995tr,Wright:1998mk}
we expect rather light sleptons.
For a gluino lighter than 1 TeV the left-handed slepton mass is
lighter than
300 GeV. 

Lepton flavor violation places a constraint
$$|(\delta^\ell_{12})_{LR}|<1.7\times 10^{-6}\left({\tilde m_\ell\over
100\ \GeV}\right)^2\ ,$$ 
whereas, using \Eref{a},  our scenario
gives
$$|(\delta^\ell_{12})_{LR}|\sim 
{A_0 m_\mu {\rm Max}(V_{12}^{\ell^-},V_{12}^{\ell^+})\over
\tim_\ell}\sim 8\times10^{-5}
\left({A_0\over 100 \GeV}\right)
\left({(100 \GeV)^2\over \tim_\ell}\right)  \ ,
$$
where in the last step we have assumed 
$${\rm Max}(V_{12}^{\ell^-},V_{12}^{\ell^+})\sim \sqrt{m_e\over
m_\mu}\ . 
$$
We therefore must require that
\bel{azero}
{A_0\over 100 \GeV}\alt 0.02  \left({\tilde m_\ell\over
100 \ \GeV}\right)^4\ .
\ee 

The constraint from the electron EDM is
\bel{edm}
{\rm Im}(\delta^\ell_{11})_{LR}<3.7\times 10^{-7}\left({\tilde m_\ell\over
100\  \GeV}\right)^2\ ,
\ee
whereas we expect
$$|(\delta^\ell_{11})_{LR}|\sim 
{A_0 m_e \over
\tim_\ell}\sim 5\times10^{-6}
\left({A_0\over 100 \GeV}\right)
\left({(100 \GeV)^2\over \tim_\ell}\right)  \ .
$$
Thus the electron EDM  places a slightly less stringent constraint on
$A_0$ than  $\mu\rightarrow e \gamma$. 
The constraint from the neutron
EDM will be even weaker, since the squarks are so much heavier
than the sleptons.
Similarly, one can look at constraints on A terms from $\tau\rightarrow
\mu\gamma$.
We have the constraint
\bel{taumugamma}
|(\delta^\ell_{23})_{LR}|<2.0\times 10^{-2}\left({\tilde m_\ell\over
100\  \GeV}\right)^2\ ,
\ee
whereas we from \Eref{a} we expect
\bel{tau}(\delta^\ell_{23})_{LR}\sim 
{A_0 m_\tau \over
\tim_\ell}\sim 1.7\times 10^{-2}
\left({A_0\over 100 \GeV}\right)\left({(100 \GeV)^2\over \tim_\ell}\right)
\ ,
\ee 
where we have assumed maximal mixing between the left-handed $\tau$
and $\mu$.
This  constraint on $A_0$ is  also less severe.

We conclude
that in this scenario the natural expectations for EDM's and the
$\mu\rightarrow e \gamma$ rate are somewhat larger than the 
experimental bounds,
unless there is some   additional   suppression of 
$A$ terms relative to gaugino masses. 

For 
squark masses of 500 GeV there is a
supersymmetric contribution to $\epsilon'/\epsilon$ of order the
experimental size from graphs containing CP-  and flavor-violating A terms.
Since the squarks and gluino must be rather heavy, and $A_0$ small,
the expectation for supersymmetric contributions to
$\epsilon'/\epsilon$
is  below the experimental value.
 Notice that we avoid the significant supersymmetric contribution to
$\epsilon'/\epsilon$ discussed in ref.~\cite{Kagan:1999iq}, 
since $\tilde {\bar u}$ and $\tilde {\bar d}$ squarks 
are degenerate to a few percent.

\subsection{$b\rightarrow s \gamma$ } 
 
The constraints from $b\to s\gamma$ are not simple 
since there may be several supersymmetric contributions, even for
flavor-degenerate scalar mass matrices. 
However, roughly speaking we must have 
$$ 
|(\delta_{23}^d)_{LR}| 
\sim {y_b\over\sqrt{4\pi}}V_{cb}  
< 10^{-2}\left({m_{\tilde d}\over 500\GeV}\right)^2 
$$ 
This is not a serious concern.

\section{Summary} 

We have shown that both the flavor hierarchy and the absence of 
unacceptable flavor changing neutral currents  from 
superpartner exchange may be explained by couplings of the first two  
generations of quark and lepton superfields to a superconformal sector at 
short distances. Using some of the recent exact results on
renormalization of soft supersymmetry breaking terms, we find a
distinctive
imprint  on the 
superpartner spectrum. The masses of 
the superpartners of the first two generations are flavor-degenerate,
and related to the gaugino masses, and the trilinear scalar terms must
be relatively small, although they will inherit the basic flavor structure
of the Yukawa couplings.  There is almost complete 
freedom for the other soft supersymmetry breaking parameters. 

While this work was in the process of completion 
reference ~\cite{Kobayashi:2001kz}
appeared, which contains some overlapping results. 

\noindent\medskip\centerline{\bf Acknowledgments} 

A.E.N. is supported in part by DOE grant \#DE-FG03-96ER40956. 
The work of M.J.S. was 
supported in part by National Science Foundation grant NSF PHY95-13835 and by 
the W.M. Keck Foundation.  
 
\bibliography{soft} 

\providecommand{\href}[2]{#2}\begingroup\raggedright\begin{thebibliography}{10}

\bibitem{Nelson:2000sn}
A.~E. Nelson and M.~J. Strassler, {\it Suppressing flavor anarchy},  {\em JHEP}
  {\bf 09} (2000) 030, [\href{http://xxx.lanl.gov/abs/hep-ph/0006251}{{\tt
  hep-ph/0006251}}].

\bibitem{Hisano:1997ua}
J.~Hisano and M.~Shifman, {\it Exact results for soft supersymmetry breaking
  parameters in supersymmetric gauge theories},  {\em Phys. Rev.} {\bf D56}
  (1997) 5475--5482, [\href{http://xxx.lanl.gov/abs/hep-ph/9705417}{{\tt
  hep-ph/9705417}}].

\bibitem{Avdeev:1997vx}
L.~V. Avdeev, D.~I. Kazakov, and I.~N. Kondrashuk, {\it Renormalizations in
  softly broken {SUSY} gauge theories},  {\em Nucl. Phys.} {\bf B510} (1998)
  289, [\href{http://xxx.lanl.gov/abs/hep-ph/9709397}{{\tt hep-ph/9709397}}].

\bibitem{Jack:1997pa}
I.~Jack and D.~R.~T. Jones, {\it The gaugino beta-function},  {\em Phys. Lett.}
  {\bf B415} (1997) 383, [\href{http://xxx.lanl.gov/abs/hep-ph/9709364}{{\tt
  hep-ph/9709364}}].

\bibitem{Jack:1998eh}
I.~Jack, D.~R.~T. Jones, and A.~Pickering, {\it Renormalisation invariance and
  the soft beta functions},  {\em Phys. Lett.} {\bf B426} (1998) 73--77,
  [\href{http://xxx.lanl.gov/abs/hep-ph/9712542}{{\tt hep-ph/9712542}}].

\bibitem{Jack:1998iy}
I.~Jack, D.~R.~T. Jones, and A.~Pickering, {\it The soft scalar mass
  beta-function},  {\em Phys. Lett.} {\bf B432} (1998) 114--119,
  [\href{http://xxx.lanl.gov/abs/hep-ph/9803405}{{\tt hep-ph/9803405}}].

\bibitem{Karch:1998qa}
A.~Karch, T.~Kobayashi, J.~Kubo, and G.~Zoupanos, {\it Infrared behavior of
  softly broken s{QCD} and its dual},  {\em Phys. Lett.} {\bf B441} (1998) 235,
  [\href{http://xxx.lanl.gov/abs/hep-th/9808178}{{\tt hep-th/9808178}}].

\bibitem{Karch:1998xt}
A.~Karch, D.~Lust, and G.~Zoupanos, {\it Superconformal {N = 1} gauge theories,
  beta-function invariants and their behavior under seiberg duality},  {\em
  Phys. Lett.} {\bf B430} (1998) 254--263,
  [\href{http://xxx.lanl.gov/abs/hep-th/9804074}{{\tt hep-th/9804074}}].

\bibitem{Kobayashi:1998jq}
T.~Kobayashi, J.~Kubo, and G.~Zoupanos, {\it Further all-loop results in
  softly-broken supersymmetric gauge theories},  {\em Phys. Lett.} {\bf B427}
  (1998) 291, [\href{http://xxx.lanl.gov/abs/hep-ph/9802267}{{\tt
  hep-ph/9802267}}].

\bibitem{Giudice:1998ni}
G.~F. Giudice and R.~Rattazzi, {\it Extracting supersymmetry-breaking effects
  from wave- function renormalization},  {\em Nucl. Phys.} {\bf B511} (1998)
  25--44, [\href{http://xxx.lanl.gov/abs/hep-ph/9706540}{{\tt
  hep-ph/9706540}}].

\bibitem{Arkani-Hamed:1998kj}
N.~Arkani-Hamed, G.~F. Giudice, M.~A. Luty, and R.~Rattazzi, {\it
  Supersymmetry-breaking loops from analytic continuation into superspace},
  {\em Phys. Rev.} {\bf D58} (1998) 115005,
  [\href{http://xxx.lanl.gov/abs/hep-ph/9803290}{{\tt hep-ph/9803290}}].

\bibitem{Arkani-Hamed:1998wc}
N.~Arkani-Hamed and R.~Rattazzi, {\it Exact results for non-holomorphic masses
  in softly broken supersymmetric gauge theories},  {\em Phys. Lett.} {\bf
  B454} (1999) 290, [\href{http://xxx.lanl.gov/abs/hep-th/9804068}{{\tt
  hep-th/9804068}}].

\bibitem{Luty:1999qc}
M.~A. Luty and R.~Rattazzi, {\it Soft supersymmetry breaking in deformed moduli
  spaces, conformal theories and {N = 2} yang-mills theory},  {\em JHEP} {\bf
  11} (1999) 001, [\href{http://xxx.lanl.gov/abs/hep-th/9908085}{{\tt
  hep-th/9908085}}].

\bibitem{Kobayashi:2000wk}
T.~Kobayashi and K.~Yoshioka, {\it New rg-invariants of soft supersymmetry
  breaking parameters},  {\em Phys. Lett.} {\bf B486} (2000) 223--227,
  [\href{http://xxx.lanl.gov/abs/hep-ph/0004175}{{\tt hep-ph/0004175}}].

\bibitem{Masiero:1999ub}
A.~Masiero and H.~Murayama, {\it Can epsilon'/epsilon be supersymmetric?},
  {\em Phys. Rev. Lett.} {\bf 83} (1999) 907--910,
  [\href{http://xxx.lanl.gov/abs/hep-ph/9903363}{{\tt hep-ph/9903363}}].

\bibitem{Cheng:1998xg}
H.-C. Cheng and Y.~Shadmi, {\it Duality in the presence of supersymmetry
  breaking},  {\em Nucl. Phys.} {\bf B531} (1998) 125--150,
  [\href{http://xxx.lanl.gov/abs/hep-th/9801146}{{\tt hep-th/9801146}}].

\bibitem{Gabbiani:1996hi}
F.~Gabbiani, E.~Gabrielli, A.~Masiero, and L.~Silvestrini, {\it A complete
  analysis of fcnc and cp constraints in general susy extensions of the
  standard model},  {\em Nucl. Phys.} {\bf B477} (1996) 321--352,
  [\href{http://xxx.lanl.gov/abs/hep-ph/9604387}{{\tt hep-ph/9604387}}].

\bibitem{Bagger:1997gg}
J.~A. Bagger, K.~T. Matchev, and R.-J. Zhang, {\it Qcd corrections to
  flavor-changing neutral currents in the supersymmetric standard model},  {\em
  Phys. Lett.} {\bf B412} (1997) 77--85,
  [\href{http://xxx.lanl.gov/abs/hep-ph/9707225}{{\tt hep-ph/9707225}}].

\bibitem{Ciuchini:1998ix}
M.~Ciuchini {\em et.~al.}, {\it Delta m(k) and epsilon(k) in susy at the
  next-to-leading order},  {\em JHEP} {\bf 10} (1998) 008,
  [\href{http://xxx.lanl.gov/abs/hep-ph/9808328}{{\tt hep-ph/9808328}}].

\bibitem{Ciuchini:1998bw}
M.~Ciuchini {\em et.~al.}, {\it Next-to-leading order qcd corrections to
  delta(f) = 2 effective hamiltonians},  {\em Nucl. Phys.} {\bf B523} (1998)
  501--525, [\href{http://xxx.lanl.gov/abs/hep-ph/9711402}{{\tt
  hep-ph/9711402}}].

\bibitem{Krauss:1998mt}
F.~Krauss and G.~Soff, {\it Next-to-leading order {QCD} corrections to b anti-b
  mixing and epsilon(k) within the mssm},
  \href{http://xxx.lanl.gov/abs/hep-ph/9807238}{{\tt hep-ph/9807238}}.

\bibitem{Contino:1998nw}
R.~Contino and I.~Scimemi, {\it The supersymmetric flavor problem for heavy
  first-two generation scalars at next-to-leading order},  {\em Eur. Phys. J.}
  {\bf C10} (1999) 347--356,
  [\href{http://xxx.lanl.gov/abs/hep-ph/9809437}{{\tt hep-ph/9809437}}].

\bibitem{Buras:2000if}
A.~J. Buras, M.~Misiak, and J.~Urban, {\it Two-loop qcd anomalous dimensions of
  flavour-changing four- quark operators within and beyond the standard model},
   {\em Nucl. Phys.} {\bf B586} (2000) 397--426,
  [\href{http://xxx.lanl.gov/abs/hep-ph/0005183}{{\tt hep-ph/0005183}}].

\bibitem{Feng:2000kg}
T.-F. Feng, X.-Q. Li, W.-G. Ma, and F.~Zhang, {\it Complete analysis on the nlo
  susy-qcd corrections to b0 - anti-b0 mixing},  {\em Phys. Rev.} {\bf D63}
  (2001) 015013, [\href{http://xxx.lanl.gov/abs/hep-ph/0008029}{{\tt
  hep-ph/0008029}}].

\bibitem{Barbieri:1988fn}
R.~Barbieri and G.~F. Giudice, {\it Upper bounds on supersymmetric particle
  masses},  {\em Nucl. Phys.} {\bf B306} (1988) 63.

\bibitem{Dimopoulos:1995mi}
S.~Dimopoulos and G.~F. Giudice, {\it Naturalness constraints in supersymmetric
  theories with nonuniversal soft terms},  {\em Phys. Lett.} {\bf B357} (1995)
  573--578, [\href{http://xxx.lanl.gov/abs/hep-ph/9507282}{{\tt
  hep-ph/9507282}}].

\bibitem{Anderson:1995tr}
G.~W. Anderson and D.~J. Castano, {\it Naturalness and superpartner masses or
  when to give up on weak scale supersymmetry},  {\em Phys. Rev.} {\bf D52}
  (1995) 1693--1700, [\href{http://xxx.lanl.gov/abs/hep-ph/9412322}{{\tt
  hep-ph/9412322}}].

\bibitem{Wright:1998mk}
D.~Wright, {\it Naturally nonminimal supersymmetry},
  \href{http://xxx.lanl.gov/abs/hep-ph/9801449}{{\tt hep-ph/9801449}}.

\bibitem{Kagan:1999iq}
A.~L. Kagan and M.~Neubert, {\it Large delta(i) = 3/2 contribution to
  epsilon'/epsilon in supersymmetry},  {\em Phys. Rev. Lett.} {\bf 83} (1999)
  4929--4932, [\href{http://xxx.lanl.gov/abs/hep-ph/9908404}{{\tt
  hep-ph/9908404}}].

\bibitem{Kobayashi:2001kz}
T.~Kobayashi and H.~Terao, {\it Sfermion masses in nelson-strassler type of
  models: Susy standard models coupled with scfts},
  \href{http://xxx.lanl.gov/abs/hep-ph/0103028}{{\tt hep-ph/0103028}}.

\bibitem{Novikov:1983uc}
V.~A. Novikov, M.~A. Shifman, A.~I. Vainshtein, and V.~I. Zakharov, {\it Exact
  gell-mann-low function of supersymmetric yang-mills theories from instanton
  calculus},  {\em Nucl. Phys.} {\bf B229} (1983) 381.

\bibitem{Novikov:1986rd}
V.~A. Novikov, M.~A. Shifman, A.~I. Vainshtein, and V.~I. Zakharov, {\it Beta
  function in supersymmetric gauge theories: Instantons versus traditional
  approach},  {\em Phys. Lett.} {\bf B166} (1986) 329--333.

\bibitem{Shifman:1986zi}
M.~A. Shifman and A.~I. Vainshtein, {\it Solution of the anomaly puzzle in
  {SUSY} gauge theories and the wilson operator expansion},  {\em Nucl. Phys.}
  {\bf B277} (1986) 456.

\bibitem{Fischler:1981zk}
W.~Fischler, H.~P. Nilles, J.~Polchinski, S.~Raby, and L.~Susskind, {\it
  Vanishing renormalization of the d term in supersymmetric u(1) theories},
  {\em Phys. Rev. Lett.} {\bf 47} (1981) 757.

\bibitem{Dine:1996ui}
M.~Dine, {\it Supersymmetry phenomenology (with a broad brush)},
  \href{http://xxx.lanl.gov/abs/hep-ph/9612389}{{\tt hep-ph/9612389}}.

\end{thebibliography}\endgroup
\bibliographystyle{JHEP} 
 
\appendix
\section{Generalization of Exact Formulas for running of Soft
Supersymmetric Terms}

In this section, we discuss some exact formulas for running of
supersymmetry-preserving and -breaking couplings.  All of this
material has appeared in the 
literature
\cite{Hisano:1997ua,Jack:1997pa,Avdeev:1997vx,Jack:1998eh,Jack:1998iy,Kobayashi:1998jq,Giudice:1998ni,Arkani-Hamed:1998kj,Arkani-Hamed:1998wc}; 
however, it has not
been well-summarized elsewhere.

As a warm-up, 
consider a  theory with four supercharges in two, three or four
dimensions.  The theory has a single chiral superfield $\phi$ and
Lagrangian
$$
\int d^4\ta  \phi^\dagger \phi + 
\left[\int d^2 \ta\ W(\phi)
+ h.c.\right]
$$
where the superpotential $W$ takes the form
$$
W(\phi) =\hat y_2 \phi^2 + \hat y_3 \phi^3 + 
\hat y_4\phi^4 + \cdots
$$
 For the moment, we limit ourselves to perturbatively renormalizable models,
implying $\hat y_k=0$ for all $k>4$ in $d=3$ and for all $k>3$ in
$d=4$.

We have defined the theory, and  normalized the kinetic
term of the scalars, at a certain scale $\mu_0$.  At this scale the
physical couplings $y_k$ are the same as the holomorphic couplings
$\hat y_k$. However, if we consider the theory at a lower scale $\mu$,
using a Wilsonian prescription, the same theory must be written
$$
\int d^4\ta  Z(\mu)\phi^\dagger \phi + 
\left[\int d^2 \ta\ (\hat y_2 \phi^2 + \hat y_3 \phi^3 + 
\hat y_4\phi^4 + \cdots) 
+ h.c.\right]
$$
where the absence of any vertex factors is a consequence of the
non-renormalization theorems.  Now the physical coupling is
related to the holomorphic one by
$$
y_k(\mu) = Z(\mu)^{-k/2} \hat y_k \equiv \sqrt {Z_{y_k}(\mu)}  \hat y_k \ .
$$
 From this we see
$$
\beta_{y_k} = {\partial y_k \over \partial \ln\mu} = {k\over 2} 
y_k\gamma_\phi \equiv 
\half y_k \gamma_{y_k} , \ {i.e.}, \ \beta_{\ln y_k} = \half \gamma_{y_k}
$$
where
$$
\gamma_\phi = - {\partial\ln Z\over \partial \ln \mu}
$$
is the anomalous mass dimension of the field $\phi$ (positive, by
unitarity, for all gauge invariant fields), and
$$
\gamma_{y_k} = - {\partial\ln Z_{y_k}\over \partial \ln \mu} = k\gamma_\phi
$$
measures the dimensionality of the coupling $y_k$.

Since $y_k$ has canonical dimension $d^0(y_k)=(d -1) - kd_\phi =
d-1-k(d-2)/2$, its beta function does not vanish at a conformal fixed
point; instead, a fixed point has
\bel{betausual}
\beta_{y_k} =  d^0(y_k)
\ee
It is often useful instead to define a dimensionless coupling
$$
{\cal Y}_k = {y_k \mu^{k({d\over2}-1)+1-d}}
$$
with
$$
\beta_{\ln {\cal Y}_k} = {\partial {\ln \cal Y}_k \over \partial \ln\mu} = 
\half \left[k(d-2)-2d+2 +{k} \gamma_\phi\right] \equiv 
\half \gamma_{{\cal Y}_k}
$$
so that a fixed point of a theory occurs at $\beta_{{\cal Y}_k}=0$.
However, we will not usually need this.

Note that these beta functions are not generally related to the
presence of  short distance
singularities.  For example, in three dimensions the theory
$$
W = y_3\phi^3
$$
is finite; but the beta function is non-zero and reaches
a fixed point in the infrared.

Note also that it is essential to treat mass thresholds in a 
particular way in order to maintain this structure through
the threshold.

\subsection{A simple example}

Now let us consider the case where only $y_3$ is non-zero; in the
following we will drop the subscript 3.  
The beta function is
$$
\beta_y =\half y[3\gamma_\phi] = 
\half y\gamma_{y} \ .
$$
Thus the theory reaches a fixed point
$\gamma_\phi=(4-d)/3$.  There can be
no such fixed point in four dimensions (since $\gamma$ must be
non-zero and positive for an interacting gauge-invariant field) but a
fixed point exists for any dimension less than four.  (This is  the
supersymmetric Wilson-Fisher fixed point.)

Next, let us consider the effect of \susy\ breaking in this theory.
At the cutoff scale, we may add a trilinear scalar term $A_0\phi^3$
and a scalar mass term $\tilde m_0^2\phi^*\phi$.  We can implement the
first term by making $y$ a superfield of the form ${\bf \hat y} =\hat
y + {\ta^2} \hat A$ .
It is
convenient to write
$$
{\bf \hat y} =\hat y e^{\ta^2 (\hat A/\hat y)} \ .
$$ 
The
scalar mass term may be added by making $Z$ a superfield of the form
${\bf Z} = Z(1-\ta^4 \tilde m_0^2)$ at some scale $\mu_0$.  However,
while ${\bf \hat y}$ is holomorphic and is not renormalized, the same
is not true for ${\bf Z}$.  At any $\mu\neq \mu_0$, we must assume it
is a general real superfield with $\ta^2$, $\bar\ta^2$ and
$\ta^4=\ta^2\bar\ta^2$ components.  It is convenient to write
$$
{\bf Z}= Ze^{\ta^2 C+\bar\ta^2 C^* -\ta^4 \DD}
$$
where $C=0$ and $\DD= \tilde m_0^2$ at $\mu=\mu_0$.

Now, the theory has a global symmetry $U(1)_\phi$ under which
\bel{glosym}
\phi \to \phi e^{\TT}, \  \phi^\dagger \to \phi^\dagger e^{\TT^\dagger}, \
{\bf \hat y}\to {\bf \hat y}e^{-3\TT},\  
{\bf Z}\rarr e^{-(\TT^\dagger+\TT)}{\bf Z}\ .
\ee
where $\TT$ is an arbitrary chiral superfield.    Note that 
this symmetry can be used to change the scale $\mu_0$
at which $C=0$ and $\DD= \tilde m_0^2$; however the transformations
are $\mu$-independent and therefore cannot fix $C=0$ at all $\mu$.
This symmetry is just
a reparametrization of our theory, and physical observables must be
invariant under such  transformations.  
The theory also has an R symmetry under which $\phi$ and $Z$ are
invariant and $\yy$ transforms with charge $2$.  Let us construct
some objects which are invariant under these symmetries.  The
simplest invariant superfield is
$$
-\bar D^2 D^2\  \ln {\bf Z} = \DD
$$
which is independent of $\ta$.  Thus $\ln\ZZ|_{\ta^4}$ is a
physically meaningful quantity, while $\ln\ZZ|_{\ta^2}$ is not.
Another important invariant is the superfield
$$
e^{\VV{y}}\equiv{\bf \hat y}^\dagger {\bf Z}^{-3}{\bf \hat y} 
= Z^{-3}|\hat y|^2 e^{
\ta^2 \left({\hat A\over \hat y} - 3C\right)+
\bar \ta^2 \left({\hat A^*\over \hat y^*} - 3C\right)
+ \ta^4 \DD_y }\ 
\ ;
$$
here $\DD_y\equiv 3 \DD$. 

These two invariants, $\tim$ and $\VV{y}$, completely specify the
running coupling constants of the theory.  (In this  case,
the D-term $\tim_y$ of $\VV{y}$ also specifies $\tim$. In some cases,
however, there are independent invariants of the form
$\bar D^2 D^2 \ \ln\ZZ$.)  Let us define
these physical parameters.  At any scale $\mu_0$, we may canonically
normalize $\phi$ by making a global {\it non-holomorphic} version of
the transformation \eref{glosym}, with $\TT= {1\over 2} \ln Z + \ta^2
C$.  After this transformation we have
$$
\yy \to Z^{-3/2}\hat y e^{\left({\hat A\over \hat y}-3C\right) \ta^2} , \
\ZZ = e^{-\ta^4 \DD}
$$
from which we conclude that the physical Yukawa coupling is
$$
y = Z^{-3/2}\hat y \ , \ \ {\cal Y} = y\mu^{d-4}
$$
(as in the supersymmetric case), the physical $A$-term is
$$
{A} = Z^{-3/2} {\hat A-3C \hat y} \ \Rightarrow \ 
{A\over y} ={\hat A\over \hat y}-3C = \VV{y}|_{\ta^2}\  ,
$$
and  the physical soft mass is
$$
\tim = -\bar D^2 D^2 \ln\ZZ = \DD \ .
$$
It is useful to define $\tim_y\equiv\DD_y = \VV{y}|_{\ta^4} = 3\tim$.

Now, this language makes the beta functions for the physical 
parameters extremely simple.  Since
$$
\VV{y} = \ln {\bf \hat y}^\dagger +  \ln {\bf \hat y} 
-3  \ln {\bf Z}
$$
and since $ {\bf \hat y}$ is $\mu$-independent by the non-renormalization
theorem of \none\ supersymmetry, we have a natural superfield beta
function
\bel{superbeta}
{\cal B}_{{\VV{y}}}\equiv \dbyd{{\VV{y}}}{\ln \mu} = -3\dbyd{\ln\ZZ}{\ln \mu} \equiv 
- \dbyd{\ln\ZZ_y}{\ln \mu}\equiv \Gamma_y\  .
\ee
where $\Gamma_y$ is a superfield anomalous dimension.  
Since ${\VV{y}} $ is invariant under the symmetries, the same must be true for
$\Gamma_y$; this follows from the fact that under \eref{glosym},
$\ln\ZZ$ shifts additively by $\TT+\TT^\dagger$, which is
$\mu$-independent.  It follows then that $\Gamma_y$ is a function only
of $\VV{y}$, $\bar D^2 D^2 \ln\ZZ$, and $\mu$.  However, since $\bar
D^2 D^2 \ln\ZZ$ vanishes in a supersymmetric theory, and since
supersymmetric formulas must be accurate when the scale of
supersymmetry breaking is small compared to the scale $\mu$, $\Gamma_y$
can only depend on positive powers of 
$(\bar D^2 D^2\ln\ZZ)/\mu^2$. For $\mu$ much larger 
than all supersymmetry-breaking
mass scales, the dependence on this linear combination can be
neglected.  In this paper, we need formulas appropriate for $M_{pl} >
\mu > M_< \gg M_W$, and so we may take $\Gamma_y$ to be a function of
$\VV{y}$ only.

All $\ta^2$ dependence therefore enters through $\VV{y}$.  The $\ta=0$
component of \Eref{superbeta} gives
$$
\beta_{\ln |y|^2} = \gamma_{y} = 2\beta_{\ln {y}} \ ,
$$
which agrees as it must with the supersymmetric case. 
The $\ta^2$ component of
\Eref{superbeta} reads
$$
\beta_{A\over y} =
\dbyd{\gamma_{y}}{\ln \VV{y}} { \ln\VV{y}}\Big|_{\ta^2} = 
{A\over y}\dbyd{\gamma_y}{\ln |y|^2} \ ,
$$
while the $\ta^4$ component gives
$$
\beta_{\tilde m^2_y}  
=3\left[\ddbyd{\gamma_y}{[\ln |y|^2]} 
\left|\left(\ln \VV{y}\Big|_{\ta^2}\right)\right|^2+
\dbyd{\gamma_{y}}{\ln |y|^2} {\ln \VV{y}}\Big|_{\ta^4}
 \right]
= \left[ \left|{A\over y}\right|^2 \ddbyd{}{[\ln |y|^2]} 
+ 3\tilde m^2 \dbyd{}{\ln |y|^2} 
\right]\gamma_{y}\ .
$$
More conventionally, we may use
$$
\dbyd{f(|y|^2)}{\ln y} = \dbyd{f(|y|^2)}{\ln y^*} = \dbyd{f(|y|^2)}{\ln |y|^2} 
$$
and write
$$
\beta_{\ln { y}} = \half\gamma_{{y}}\ ,
$$
$$
\beta_{A\over y} = 
{A\over y}\dbyd{}{\ln y}\gamma_y \equiv -D_1\gamma_y\ ,
$$
and
$$
\beta_{\tilde m^2_y} = \left[
\left|{A\over y}\right|^2 \ddbydd{}{\ln y}{\ln y^*}
+\tilde m^2_y \dbyd{}{\ln y}
\right]\gamma_{\phi} \equiv \left[D_1^*D_1 + D_2\right]\gamma_{y}\ ,
$$
or equivalently, dividing by $3$,
$$
\beta_{\tilde m^2} =  \left[D_1^*D_1 + D_2\right]\gamma_{\phi}\ .
$$
where
$$
D_1 \equiv -{A\over y}\dbyd{}{\ln y} \ , \ 
D_2\equiv \tilde m^2_y \dbyd{}{\ln y} \ .
$$
Note that our formal definition of $D_1^*D_1$ requires that derivatives
with respect to $\ln y^*$ do not act on $A\over y$.

\subsection{More couplings}

Still working with one field $\phi$, let us allow $N_y$ of the
renormalizable $\hat y_k$ to be non-zero.  Let us call $N_I$ the
number of independent invariants $\VV{m}$.  It is easy to see that there
are many more invariants than there are couplings, since
$\yy_k^\dagger\ZZ^{-2k}\yy_k\equiv\exp{\VV{k}}$ is always present, but
for example $\yy_4^\dagger\yy_5^\dagger\ZZ^{-9}\yy_3\yy_6$ is an
additional invariant under $U(1)_\phi$ and $U(1)_R$. In general, every
coupling invariant will have the form
\bel{invdef}
\VV{m} = \sum_k \QQ_{mk} \ln\yy_k + \sum_k  \bQQ_{mk} \ln\yy^\dagger_k 
- (\sum_k\QQ_{mk}) \ln\ZZ 
\ee
where $\QQ_{mk}$ and $\bQQ_{mk}$ are two independent $N_I\times N_y$
matrices, with all entries positive or zero, satisfying the two constraints
\bel{QR}
\sum_k\QQ_{mk} = \sum_k\bQQ_{mk}\ \  (U(1)_R \ {\rm conservation}),
\ee
\bel{Qphi}
\sum_k k\QQ_{mk} = \sum_k k \bQQ_{mk}\ \ (U(1)_\phi \ {\rm conservation}).
\ee

The components of \Eref{invdef} are, in terms of the physical
parameters,
$$
\VV{m}\Big|_{\ta=0} = \sum_k \QQ_{mk} \ln y_k + \sum_k \bQQ_{mk} \ln y^*_k 
$$
$$
\VV{m}\Big|_{\ta^2} =\sum_k  \QQ_{mk} {A_k\over y_k} \ , \ 
\VV{m}\Big|_{\bar\ta^2} = \sum_k \bQQ_{mk} {A^*_k\over y^*_k} 
\neq (\VV{m}\Big|_{\ta^2})^*
$$
$$
\VV{m}\Big|_{\ta^4} = \sum_k \QQ_{mk} \tim_k = \sum_k \bQQ_{mk}\tim_k
$$
where the fact that $\tim_k = k\tim$, combined with the $U(1)_\phi$
constraint \Eref{Qphi}, implies the two expressions for the $\ta^4$
component are equal.  From the components of 
$\VV{k}$, $k=1,\dots,N_y$, we confirm that the beta functions
for the couplings $y_k$ satisfy
$$
\beta_{\ln y_k} = \half {\gamma_k} \ , \
\beta_{{A_k\over y_k}} = -D_1 \gamma_k \ , \ 
\beta_{{A_k^*\over y_k^*}} = -\bar D_1 \gamma_k \ , \
\beta_{\tim_k} = (\bar D_1 D_1 + D_2 )\gamma_k
$$
where $\gamma_k = k\gamma_\phi$,
$$
D_1 = -\sum_m \VV{m}\Big|_{\ta^2} \dbyd{}{\VV{m}}\ , \ 
\bar D_1 = -\sum_m \VV{m}\Big|_{\bar\ta^2} \dbyd{}{\VV{m}} \ ,
$$
(note $\bar D_1$ is not apparently the complex conjugate of $D_1$,
since not all $\VV{m}$ are real) and
$$
D_2 = \sum_m \VV{m}\Big|_{\ta^4} \dbyd{}{\VV{m}} \ .
$$
But since, acting on any function of the invariants,
$$
\dbyd{}{\ln y_k} = \sum_m \QQ_{mk} \dbyd{}{\VV{m}} \ , \
\dbyd{}{\ln y_k^*} = \sum_m \bQQ_{mk} \dbyd{}{\VV{m}} \ ,
$$
we have
$$
D_1 = -\sum_{k,m} {A_k\over y_k}\QQ_{mk} \dbyd{}{\VV{m}} 
=-\sum_k {A_k\over y_k}\dbyd{}{\ln y_k}  \ ,
$$
$$
\bar D_1 = -\sum_m \VV{m}\Big|_{\bar\ta^2} \dbyd{}{\VV{m}} =
-\sum_j {A^*_j\over y^*_j} \dbyd{}{\ln y^*_j} = (D_1)^* \ ,
$$
and
$$
\sum_m \VV{m}\Big|_{\ta^4} \dbyd{}{\VV{m}} = 
\sum_k  \tim_k \dbyd{}{\ln y_k} = 
\sum_k  \tim_k \dbyd{}{\ln y_k^*}
$$
which implies
$$
D_2= \sum_j \tim_j \dbyd{}{\ln y_j} = \half\tim_j \sum_j
\left(\dbyd{}{\ln y_j} +\dbyd{}{\ln y_j^* }\right) \ .
$$
So in the end we need not worry about the particular
invariants in a given theory; $D_1$ and $D_2$ only
involve derivatives with respect to the $N_y$ physical couplings.

\subsection{More Fields}
At this point it is straightforward to generalize to a theory with 
more fields $\phi_i$ with superpotential 
\bel{PPsup}
W = \sum_{s=1}^{N_y} \  \yy_s \prod_{i} (\phi_i)^{{\cal P}^i_s} \ ,
\ee
where ${\cal P}^i_s$ is a matrix with non-negative integer entries, and
$\yy_s= \hat y_s + \theta^2 \hat A_s$.  
If we assume that  symmetries prevent mixing among the different $\phi$'s,
there are $N_y$ invariants
of the form
$$
e^{\VV{s}} = \yy_s^\dagger \yy_s
\left(\prod_i\ZZ_i^{{\cal P}^i_s}\right)^{-1}  \ \ (s=1,\dots,N_y)
$$
where $\ZZ_i$ is the wave-function renormalization of $\phi_i$.
In addition there are  invariants $\VV{m}$, $m=N_y+1,\dots,N_I$.
As before, we have
$$
\beta_{\ln y_s} = \half {\gamma_s} \ , \
\beta_{{A_s\over y_s}} = -D_1 \gamma_s \ , \ 
\beta_{{A_s^*\over y_s^*}} = -\bar D_1 \gamma_s \ , \
\beta_{\tim_s} = (\bar D_1 D_1 + D_2 )\gamma_s
$$
where 
$$
\gamma_s = 
\sum_i {\cal P}^i_s \gamma_{\phi_{i}}\ ,
$$
\bel{msdef}
\tim_s\equiv {\cal P}^i_s \tim_i \ .
\ee
and where
$$
D_1 = -\sum_{m=1}^{N_I} \VV{m}\Big|_{\ta^2} \dbyd{}{\VV{m}}
=-\sum_s {A_s\over y_s}\dbyd{}{\ln y_s}  
\ , \ 
\bar D_1 
= (D_1)^* \ ,
$$
and
$$
D_2 = \sum_{m=1}^{N_I} \VV{m}\Big|_{\ta^4} \dbyd{}{\VV{m}} 
= \half\tim_s \sum_s
\left(\dbyd{}{\ln y_s} +\dbyd{}{\ln y_s^* }\right)\ . 
$$
The proofs of these statements are essentially identical to those
given above.

\subsection{Mixing}
\def\llambda{y}
\def\yaa{{{\bf \llambda}_s}_{ a_1 . . . a_n}}
\def\ybb{{{\bf \llambda}_s}_{ b_1 . . . b_n}}
\def\yaas{{{\bf \llambda}_s}^{* a_1 . . . a_n}}
\def\ybbs{{{\bf \llambda}_s}^{* b_1 . . . b_n}}
\def\yaahat{{{\hat \llambda}_s}{}_{ a_1 . . . a_n}}
\def\ybbhat{{{\hat \llambda}_s}{}_{ b_1 . . . b_n}}
\def\yaashat{{{\hat \llambda}_s}{}^{* a_1 . . . a_n}}
\def\ybbshat{{{\hat \llambda}_s}{}^{* b_1 . . . b_n}}
\def\yaat{{{\tilde \llambda}_s}{}_{ a_1 . . . a_n}}
\def\ybbt{{{\tilde \llambda}_s}{}_{ b_1 . . . b_n}}
\def\yaast{{{\tilde \llambda}_s}{}^{* a_1 . . . a_n}}
\def\ybbst{{{\tilde \llambda}_s}{}^{* b_1 . . . b_n}}
\def\JJ{{\bf J}}
If we have $n$ fields  which are
not distinguished by any symmetry, we must extend the  requirement of 
reparametrization invariance to include $U(n)$ transformations
\bel{nonabel}
\phi\rightarrow e^\TT \phi\ ,\ \phi^\dagger\rightarrow \phi^\dagger
e^{\TT^\dagger}\ , \ \ZZ\rightarrow e^{-\TT^\dagger}\ZZ e^{-\TT}\ ,
\ee
where now $\phi$ is an $n$ component vector, and $\TT$ and $\ZZ$ are $n$ by $n$ matrix
valued  superfields.  Superpotential couplings involving $\phi$ fields
transform like the appropriate tensors.

To simplify the construction of invariants under the transformation 
\eref{nonabel}  it is convenient to  define  superfields $\XX$ such that
\bel{ambig}\ZZ=\XX^\dagger\XX\ \ .
\ee
Under \Eref{nonabel},
$$\XX\rightarrow \XX e^{-\TT}\ .$$

Note that  there is an ambiguity in the definition of $\XX$ since we
could always redefine
\bel{unit}
\XX\rightarrow U \XX
\ee where 
$$U^\dagger U=1\ .$$
Thus physically meaningful
equations must be covariant under the transformation \eref{unit}, as
well as \eref{nonabel}.

We can now define a    superfield
\bel{Gamma}
\Gamma\equiv \left(\XX{d\XX^{-1}\over
d\ln\mu}+{d\XX^{\dagger-1}
\over d\ln\mu} \XX^\dagger\right)
\ee 
which  under \Eref{unit} transforms as
$$\Gamma\rightarrow U\Gamma U^{\dagger}\ ,$$
and is invariant under the transformation \Eref{nonabel}.
Provided that we choose at all values of $\mu$
\bel{canonical0}\XX|_{\theta=0}=\XX^\dagger|_{\theta=0}\ee
then
\bel{gamma}
\gamma=\Gamma|_{\theta=0} \ee
 is the usual anomalous dimension matrix. We refer to
$\Gamma$ as the  anomalous dimension superfield.

If we write
$$\ZZ=e^{\JJ^\dagger} e^{-\theta^4 \tim} e^{\JJ}\ ,$$ where $\JJ$ is an
ordinary chiral superfield, then we can eliminate $\JJ$ by
making a transformation \eref{nonabel}
with
$$\TT=\JJ\ .$$ Thus $\tim$ is the  soft supersymmetry breaking
mass squared matrix in a canonical basis.
By choosing 
\bel{canonical}\XX=e^{-\theta^4 \tim/2} e^{\JJ}\ ,\ee
with
\bel{canonicalII}\JJ|_{\theta=0}=\JJ^\dagger|_{\theta=0}\ ,\ee
 and examining \Eref{Gamma}, we see that the
 soft mass term beta functions can be read
off from
\bel{mbeta}
{d \tim\over d \ln \mu} = \Gamma|_{\theta^4}\ . \ee
To find the  $\theta^4$ component of $\Gamma$, we
define couplings
\bel{ydef}\ybb=\yaahat
(\XX^{-1})^{a_1}_{b_1}. . .(\XX^{-1})^{a_n}_{b_n}\ .\ee
Here $\yaahat$ is the superpotential coefficient of $\phi_s\phi_{a_1}
...\phi_{a_n}$. Unlike the $\hat y$, the ${\bf y}$ couplings are not
chiral superfields. However they are invariant under  the
transformation \eref{nonabel}. Furthermore they transform covariantly under
\eref{unit}, and 
  $\Gamma$ is a covariant function of these
 couplings  under  the transformation \Eref{unit}. 
Provided that we make the  choices eqs. \eref{canonical} and
\eref{canonicalII}, then
$$\tilde \llambda= {\bf y}|_{\theta=0}\ $$ is a superpotential
coupling in a basis with canonically normalized kinetic terms and
$$\tilde A= {\bf y}|_{\theta^2}\ $$ is the associated  
trilinear scalar coupling.
To find the $\tim$ beta function, we need to read off the
 $\theta^4$ component of $\Gamma$  in a canonical basis.

To do this, we can  define 
$$D_1\equiv -\sum_s \tilde A_s\byd{\tilde\llambda_s}$$
 as before and $D_2$ as
\bel{d2}
D_2\equiv\sum_s {1\over 2}((\tim)^{a_1}_{b_1}\delta^{a_2}_{b_2}...
\delta^{a_n}_{b_n}+{\rm perms\ of\ }a_i,b_i) 
\left(
\yaat{\partial\over \partial \ybbt}+
\ybbst{\partial\over \partial \yaast}\right)\ .
\ee
Then we have the  matrix equation
$$(\bar D_1 D_1 +D_2)\gamma ={d \tim \over d\ln\mu}\ .$$

\subsection{Four Dimensional Gauge Couplings}

Now let us turn our attention to a four-dimensional theory with a single gauge
coupling, in order to focus on some of the special issues
associated with this coupling.  Let us consider a simple group
$G$ coupled to a single matter field $\phi$ in some representation
$r$ of $G$, with index equal to twice $T_r$.  The gauge field 
strength is contained in a superfield 
$$
W_\al = -{1\over 4}\bar D\bar D e^{-V}D_\alpha e^{V}
$$
and the classical Lagrangian is given by
$$
\int d^4\ta\ \phi^\dagger e^V\phi + 
\int d^2\ta\ \SS_0\ \tr\ W_\al W^\al + {\rm hermitean \ conjugate}
$$
where the chiral background superfield
$$
\SS_0 = {1\over 2 g_0^2} (1 - 2M_0\ta^2) - {i \Theta\over 16 \pi^2}
$$
contains the bare gauge coupling $g_0$, the bare gaugino mass $M_0$,
and the theta angle. Unlike the other holomorphic couplings,
$\SS_0$ is renormalized at one loop
$$
\SS(\mu) = \SS_0 + {b_0\over 16\pi^2} \log{\mu\over \mu_0}
$$
where $\mu_0$ is the scale at which $g_0$ is defined, and $b_0 =
3T_G-T_r$ is the coefficient of the one-loop beta function.  Letting
$$
\hat\Lambda^{b_0} \equiv \mu_0^{b_0} e^{-{8\pi^2\over g_0^2}}
$$
generalize to a superfield
$$
\LLam^{b_0}\equiv \mu_0^{b_0} e^{-{16\pi^2 \SS_0}} = 
\mu^{b_0} e^{-{16\pi^2 \SS}} 
$$ 
we have
$$
\SS(\mu) = {b_0\over 16\pi^2} \log{\mu\over \LLam}
$$
Note
$$
\SS(\mu)|_{\ta^2} = \SS_0|_{\ta^2}
$$
so the running holomorphic gaugino mass divided by the running holomorphic
gauge coupling is a renormalization group invariant.

Under the $U(1)_\phi$ symmetry \eref{glosym}, $\SS$ transforms
anomalously,
\bel{glosymB}
\SS_0 \to \SS_0\ , \ \SS\to \SS - {T_r\over 8\pi^2} \TT\ , \
\LLam^{b_0}\to \LLam^{b_0} e^{2T_r \TT}
\ee
An invariant under this symmetry is
$$
\mu^{2b_0} e^{-16 \pi^2\RR_0}= 
\left[\LLam^{b_0}\right]^\dagger
\ZZ^{-2T_r}\LLam^{b_0} \ .
$$
However, the logarithm of this invariant is not the physical gauge
coupling, as it has the wrong two-loop beta function.  A suitable
coupling may defined using the NSVZ scheme \cite{Novikov:1983uc,Novikov:1986rd,Shifman:1986zi}.  NSVZ define a
different invariant, $\RR$, through the instanton factor
$$
\mu^{2b_0} e^{-16 \pi^2\RR}= 
\left[\LLam^{b_0}\right]^\dagger
\ZZ^{-2T_r}\ZZ_\lambda^{2T_G}\LLam^{b_0} \ .
$$
where $\ZZ_\lambda$ is the gaugino wave-function renormalization.
(The reason $\ZZ_\lambda$ appears with a positive sign is that
the gaugino zero modes come with a negative sign but the
more numerous gauge-boson zero modes more than compensate for this.)
Taking the logarithm we obtain
\bel{RRdef}
\RR  = \SS + \SS^\dagger 
 - {1\over 8\pi^2} 
\left( T_r \ln \ZZ - T_G \ln \ZZ_\lambda\right)
\ee
Using
$$
\ZZ_\lambda = \RR
$$
which follows from the fact that the gauge coupling and the 
wave function renormalization of the gaugino must be equal by
supersymmetry, we may also write this as
\bel{RRSS}
\RR - {T_G\over 8\pi^2} \ln \RR =  \SS + \SS^\dagger -
{1\over 8\pi^2} 
T_r \ln \ZZ = \RR_0
\ee
We will soon need the fact that 
\bel{RRderiv}
\dbyd{\RR_0}{\ln\mu} = 
\left(1- {T_G\over 8\pi^2\RR}\right)\dbyd{\RR}{\ln\mu} = 
{1\over 8\pi^2}\left(-T_r  \dbyd{\ln \ZZ}{\ln\mu}\right)
\ee

Let us now define some natural objects using these invariants.
We may write
$$
4\pi\RR = \alpha^{-1} (1 - M \ta^2 - M^* \bar \ta^2 - P\ta^4)
$$
where $\alpha=g^2/4\pi$ is the physical gauge coupling and $M$ is the physical
gaugino mass; $P$, also invariant, is not an independent quantity as
we will see in a moment.  
The gauge coupling
$$
\alpha^{-1} = 4\pi\RR|_{\ta=0}
$$
has beta function 
$$
\betau = 4\pi\dbyd{\RR}{\ln\mu}\Big|_{\ta=0}
$$
which is related to the anomalous dimension of the gluino
$$
\gamma_\lambda \equiv -\dbyd{ \ln\RR}{\ln\mu}\Big|_{\ta=0}
$$
by 
$$
\gamma_\lambda = - \alpha{\betau} =  
\beta_{\ln \alpha} =2 {\beta_\lambda\over g} = {\beta_\alpha\over \alpha}.
$$
{}From \Eref{RRSS} and \Eref{RRderiv} we obtain
$$
\gamma_\lambda =  -{b_0 + T_r \gamma_\phi \over {2\pi\over \alpha} - T_G }  \ .
$$

The gaugino mass satisfies
\bel{MdefB}
M = -{\ln \RR}\Big|_{\ta^2}
\ee
from which we learn
$$
\beta_M = -\dbyd{\ln \RR}{\ln\mu}\Big|_{\ta^2} = \dbyd{\gamma_\lambda}{\RR}\
\RR|_{\ta^2} = -{M\over \alpha} \dbyd{\gamma_\lambda}{\alpha^{-1}}\equiv D_1\gamma_\lambda
$$
where in the last steps we have used the fact that $\RR$ is an
invariant and can only depend on invariant quantities.\footnote{Alternatively
we may observe that
\bel{MdefA}
{M\over \alpha} = - 4\pi\RR|_{\ta^2} =  {1\over 2\pi} 
\left(M_0 b_0 \log {\mu_0\over \Lambda} 
+ T_r \ln\ZZ|_{\ta^2} + T_G M\right)
\ee
which gives
$$
\dbyd{}{\ln\mu}\left[M\left({2 \pi\over \alpha}- T_G\right)\right] = 
\beta_M\left[{2\pi\over \alpha}- T_G\right] + 2 \pi M \betau
=  T_r \dbyd{\ln\ZZ}{\ln\mu}\Big|_{\ta^2} = -T_rD_1\gamma_\phi \ .
$$
Since
$$
-T_r D_1\gamma_\phi 
= 
D_1 \left(\gamma_\lambda\left[{2\pi\over\alpha} - T_G\right]\right)
 =
\left[{2\pi\over \alpha} - T_G\right]
D_1 \gamma_\lambda + {2\pi M\over \alpha}\gamma_\lambda
$$
we see again that $\beta_M=D_1\gamma_\lambda$.
}
Note that if this is the only coupling, a solution to this equation is 
$M \propto \gamma_\lambda$.

Finally, \Eref{RRdef} shows
$$
{P\over \alpha} = -4\pi\RR|_{\ta^4} = 
{1\over 2\pi}(T_r \ln\ZZ - T_G \ln\RR)|_{\ta^4} = 
{1\over 2\pi}(-T_r\tim - T_G [-P - |M|^2])
$$
or
\bel{mlambdadef}
P = {  T_G |M|^2 - T_r \tim \over {2\pi\over \alpha} - T_G} \equiv 
-\alpha\tim_\lambda  \ .
\ee
Now, again using the fact that $\RR$ is an invariant and can only
depend on invariants,
$$
\beta_{\tim} = -\dbyd{\ln\ZZ}{\ln\mu}\Big|_{\ta^4} =  
\left|{\RR}|_{\ta^2}\dbyd{\gamma_\phi}{\RR}\right|^2
+\RR|_{\ta^4}\dbyd{\gamma_\phi}{\RR}
\equiv (\bar D_1 D_1 + D_2)\gamma_\phi
$$
where $D_1$ is as above, $\bar D_1= D_1^*$, and
$$
D_1 = -{M\over \alpha} \dbyd{}{\alpha^{-1}}\ , \
 \bar D_1= D_1^* \ , \ 
D_2 = \tim_\lambda\dbyd{}{\alpha^{-1}}
$$
Thus
$$
\beta_{\tim} = \left[\left|{M\over \alpha}\right|^2 \ddbyd{}{(\alpha^{-1})}   
-{T_G |M|^2   - T_r \tim \over 8\pi^2
\left(1 - {T_G \alpha\over 2 \pi}\right)}\dbyd{}{(\alpha^{-1})}\right]\gamma_\phi
$$
in agreement with the literature.

\subsection{General Four-Dimensional Theory}

It is clear now how to write the beta functions for physical
parameters in a general theory.  A general theory has invariants
$\II_m$, of the form
$$
\ln \II_m = 
\ln \sigma_m + \ln \sigma_m^* 
+ \ta^2 \omega_m + \bar\ta^2 \omega_m^* + \ta^4 \nu_m
$$
If
$${\cal B}_{\ln\II_m} = \Gamma_m$$
then
\bel{allbetas}
\beta_{\ln \sigma_m} = \half\gamma_m
\ , \
\beta_{\omega_m} = -D_1\gamma_m
\ , \
\beta_{\nu_m} = (D_1^*D_1 + D_2)\gamma_m
\ee
where
\bel{allDs}
-D_1 = \omega_n\dbyd{}{ \ln \sigma_n} \ , \
D_2 = \nu_n\dbyd{}{\ln \sigma_n} = \half\nu_n\left(\dbyd{}{\ln \sigma_n}+ \dbyd{}{\ln \sigma_n^*} 
\right)
\ee

As a check, let us verify that we obtain the results previously
in the literature for the case of a single gauge coupling
and a single Yukawa coupling.  We find
$$
D_1^*D_1 = \left|{M\over \al}\right|^2 \ddbyd{}{(1/\al)} 
+ {M^*\over \al}{A\over y}\ddbydd{}{(1/\al)}{\ln y}
+ {M\over \al}{A^*\over y^*}\ddbydd{}{(1/\al)}{\ln y^*}
+ \left|{A\over y}\right|^2\ddbydd{}{\ln y}{\ln y^*}
$$
$$
=\left|{M\al}\right|^2 
\left(\ddbyd{}{\al} + {2\over \al}\dbyd{}{\al}\right) 
- {M^* \al}{A\over y}\ddbydd{}{\al}{\ln y}
- {M \al}{A^*\over y^*}\ddbydd{}{\al}{\ln y^*}
+ \left|{A\over y}\right|^2\ddbydd{}{\ln y}{\ln y^*}
$$
while 
$$
D_2 = -\al^2 \tim_g\dbyd{}{\al}  
+ \half\tim_y \left(\dbyd{}{\ln y}+\dbyd{}{\ln y^*}\right)
={\al^2\over 2\pi}
{  T_G |M|^2 - T_r \tim \over 1- {T_G\al\over 2\pi}} \dbyd{}{\al}
+ \half\tim_y \left(\dbyd{}{\ln y}+\dbyd{}{\ln y^*}\right) \ .
$$

\section{Stability of supersymmetric fixed points}

We now prove that at a non-trivial conformal fixed point, all
A terms and gaugino masses associated with fixed non-zero couplings
and all combinations of scalar masses $\tilde m_s^2\equiv
\sum_i {\cal P}^i_s\tilde m_i^2$ run to zero.

A conformal fixed point is characterized by some non-zero
couplings $\sigma_r=\sigma_{r*}$ satisfying the conditions
$$
\beta_{\ln\sigma_r}(\sigma_v=\sigma_{v*}) = d^0_{{\sigma}_r}
$$
where $d^0_\sigma$ is the engineering dimension of
$\sigma$ [see  \Eref{betausual} and surrounding discussion.]  More generally,
near such a fixed point each beta-function superfield satisfies
$$
{\cal B}_r =d^0_{{\bf I}_r}
 + \left(\dbyd{{\cal B}_r}{\ln {\bf I}_s}\right)_*
\left(\ln {\bf I}_s-\ln {\bf I}_{s*}\right)  + \cdots
$$
where a subscript $*$ implies that the quantity is to be evaluated
{\it at} the fixed point.  Since the fixed point is supersymmetric,
\bel{gammapdef}
\Gammaprime_{rs}\equiv
\left(\dbyd{{\cal B}_r}{\ln {\bf I}_s}\right)\Bigg|_{
\ln {\bf I}_v=\ln \sigma_{v*}} =
\left(\dbyd{{\gamma_r}}{{\ \ln \sigma}_s}\right)_*
\ee
is a simple matrix of numbers.   (Note $\sigma_{v*}$ is  
a constant times the simple power of the renormalization scale 
$\mu$ needed to make it dimensionless; this $\mu^{-d^0_s}$ 
dependence does not appear
in the derivative with respect to $\ln \sigma_s$.) 

We are only interested in stable fixed points --- that is, fixed points
with the property that for all nonzero $\sigma_r$,  any 
nearby theory with $\ln\sigma_r= \ln\sigma_{r*}+\delta_r$
flows back to the original fixed point.
This implies
$$
\sum_r \delta_r \  
\beta_{\ln \sigma_r}\Bigg|_{\ln\sigma_v = \ln\sigma_{v*}+\delta_v}
>0
$$
which requires, since the $\delta_r$ are arbitrary, that
$$
\dbyd{\beta_{\ln\sigma_r}}{\ln\sigma_s}\Bigg|_{\sigma_v=\sigma_{v*}}
$$
be a matrix whose eigenvalues are all positive definite, or equivalently that $
\Gammaprime $ defined in \Eref{gammapdef} be a positive matrix.

However, the same matrix $\Gammaprime$ controls the
running of the A-terms and gaugino masses through \Eref{allbetas}-\eref{allDs}.
In particular, $\beta_{\omega_s} = -D_1\gamma_s$, which contains
$\Gammaprime$.  The positivity of $\Gammaprime$ required for
stability thus drives the A-terms and gaugino masses to zero.
Similarly, once the $\omega_s$ have run small, the appearance
of $\Gammaprime$ in $D_2\gamma_s$ ensures the linear combinations
of soft masses appearing in the $\nu_s$ also run small.

This can be stated in superfield language.  The stability of 
the fixed point requires that $\II_s \rightarrow \II_{s*}$ in the
infrared.  Since $\II_{s*}$ has no $\theta^2$ or $\theta^4$ terms,
any supersymmetry violation in $\II_s$, through $\omega_s$ and $\nu_s$,
must approach zero in the infrared\footnote{Nota Bene:
these equations do not imply that the $\omega_s$ actually {\it reach}
zero.  Although $\omega_s$ has positive anomalous dimension,
it need not have dimension greater than one. This means that at some
scale $\mu$ and $\omega_s$ are of the same order; below this scale
the above equations are no longer valid.  In our models this issue
is inconsequential, since these equations are only used at scales far higher
than 1 TeV.}.  

Notice however that the running of the $\nu_s$ to zero does not
necessarily guarantee that all soft scalar masses run toward zero.
Recall
that $\nu_s$ refers to the soft masses associated to {\it coupling
constants}; in particular for each Yukawa coupling $y_s$ the
combination $\tim_s$ defined in \Eref{msdef} runs to zero, while for gauge
couplings the combination $\tim_\lambda$ defined in \Eref{mlambdadef} runs to zero.
Given a scalar field $\phi_i$, its soft mass $\tim_i$ runs to zero
only if $\tim_i$ can be written as a linear combination of the
$\tim_s$ and $\tim_\lambda$.  This will be an essential constraint on model
building. The following simple counting argument shows that this
constraint is satisfied provided the 
strong couplings of the theory break all global non-R U(1) symmetries.

In general,  with $n$ irreducible multiplets of chiral superfields
(irreducible under nonabelian gauge or
global symmetries) of chiral superfields, we have $n$
independent soft scalar mass squared terms, and $n$  possible
independent non-R $U(1)$ global symmetries. We will now show that a
theory which explicitly breaks all these   symmetries via gauge
anomalies or superpotential terms will drive  all soft
SUSY-breaking scalar mass terms  to zero.

{}The linear combination of soft scalar masses in   \Eref{mlambdadef},
which run to zero if the associated gauge coupling is at a stable
fixed point, is precisely the same as
the linear combination of the corresponding $U(1)$'s which is
anomalous under the corresponding gauge transformation.  Furthermore, each
superpotential coupling is also associated through \Eref{msdef}  with a
linear combination of scalar masses-squared, which is again exactly the linear
combination which is explicitly broken by the corresponding
superpotential coupling. If all $n$ possible non-R $U(1)$ symmetries
are explicitly broken, by gauge anomalies and superpotential
couplings, there must be at least $n$ couplings which break independent
combinations of the possible symmetries. Then  there
must be $n$ independent linear combinations of the  scalar
masses-squared which run to zero. This is only possible if all scalar
masses-squared run to zero.

An even simpler argument shows that any exact non-R U(1) symmetry of
the strong couplings corresponds to a linear combination of scalar
masses which do not run at the fixed point.  Any such global 
symmetries could be gauged with arbitrarily weak coupling, provided
spectator chiral superfields are added to cancel anomalies.  Now any
soft SUSY-breaking mass squared terms proportional to the U(1) charge
can be written as a supersymmetric Fayet-Iliopoulos term, plus soft
terms for the spectators and a SUSY-breaking constant added to the
Hamiltonian. We have already argued that at the fixed point the
SUSY-breaking gaugino masses, A terms, and scalar masses not
proportional to  global symmetry charges  run to zero.  Thus, up to
spectator fields, and constant terms in the Hamiltonian, at the fixed
point the theory with such soft breaking terms is equivalent to a
supersymmetric theory with Fayet-Iliopoulos terms and arbitrarily weak
U(1) gauge couplings.  Fayet-Iliopoulos terms are known not to get
renormalized \cite{Fischler:1981zk,Dine:1996ui}.

\section{Flow of Supersymmetry Breaking Terms 
for the MSSM Coupled to a Conformal Field Theory}

We now consider the effect on the soft parameters of the MSSM when we 
couple its fields to those of a conformal sector.  Specifically, we
study the energy regime near and below $M_>$ and above $M_<$.  We will
generally assume that at the scale $M_<$ the trilinear scalar terms
are of order $A_0$, the gaugino masses are of order $m_{1/2}$, and the
scalar masses-squared are of order $m_0^2$; for simplicity of
presentation we will take these parameters to be of similar size, of
order 
$M_{SUSY}$.  We will show that
that\footnote{In this list we crudely separate the first two
generations from the third; however, as discussed in
\cite{Nelson:2000sn} we may well expect part of the third generation
to be strongly coupled to the conformal sector and to therefore be
more characteristic of the first two generations.}
\begin{itemize}
\item The gaugino masses associated with the conformal-sector
gauge group and  and the A-terms with the strong superpotential
couplings involving the conformal
sector and its couplings to the standard model fields are driven
down to a size  $\alpha M_{SUSY}/4\pi$, where $\alpha$ is a 
standard model gauge coupling.  These are then too
small to have a large impact on the standard model sector.
\item The third generation of the standard model, although it
retains large Yukawa couplings to the Higgs boson and corresponding
large A-terms, does not contribute dominantly to
the running of the other soft-breaking
parameters, since it decouples from the strong sector except through
two-loop effects.
\item The standard model A-terms largely
run by the same factor as their corresponding Yukawa couplings, and the
gaugino masses tend to run by a similar factor as the 
the gauge couplings $\alpha$.  
\item The soft masses of the third generation do not run by large factors.
However, the soft masses of the first two generations, which are
governed by the conformal sector, are driven to values of order 
$\tim \sim \alpha |M|^2/4\pi$, where $M$ is the gluino mass.  These values
would be zero were it not for the standard model couplings.
\end{itemize}
Thus, we claim that the after the conformal sector decouples, the
third generation soft parameters (more precisely, the top-quark
and Higgs parameters and some, or possibly all, of the bottom and tau
parameters) are at a scale not too far from $M_{SUSY}$; the gaugino masses
lie only a little below this scale; the matrices of A-terms have
roughly inherited the hierarchical structure of the Yukawa couplings
with which they are associated, suppressing their effects on
flavor-changing neutral currents; and the soft masses of the first two
generations of sfermions are of order $\tim \sim \alpha |M|^2/4\pi$,
small enough that the non-degeneracies among them will not lead to
large flavor-changing neutral currents after the additive
renormalization of \Eref{softendpt}.

Our approximate conformal fixed point has three different types of
operators which we must consider.  First, there are irrelevant
operators which control the flow into the fixed point; these are
characterized by positive eigenvalues in the matrix $\Gammaprime$,
defined in \Eref{gammapdef}.  Second, there are couplings which run
slowly, such as the standard model gauge couplings, or the top quark
Yukawa coupling.  These are characterized by eigenvalues in the matrix
$\Gammaprime$ which are very small.  We will refer to these as weak
 marginal couplings, though they are only approximately marginal.
Finally, there is at least one relevant operator whose eigenvalue in
$\Gammaprime$ is large and negative, but whose associated coupling is
very small initially.  It is this operator which drives the theory
away from the CFT when its coupling grows sufficiently large.

In the previous section we showed that if the relevant and
nearly marginal operators are set to zero, so that $\Gammaprime$ for
the nonzero couplings is a positive matrix, the approach to
the fixed point drives all irrelevant couplings and their
supersymmetric partners to zero.   The flow is characterized
by the eigenvalues of $\Gammaprime$, with the smallest eigenvalue,
or equivalently the least irrelevant operator,
determining the flow at low energy.  

In particular, we showed that for the nonzero couplings
$\sigma_{m*}$ in the strong sector, we have
\bel{vecblfpB}
 \dot{(\ln { \sigma})_r} - 
(\Gammaprime)_{rs}\ln{\sigma}_s
= {\rm order}[(\sigma-\sigma_{*})^2]
\ee
for the couplings $\sigma_r$ themselves,
\bel{vecbkB}
 \dot{{\omega}_r} -(\Gammaprime)_{rs}{\omega}_s = 0
\ee
for $\omega_r$, the associated strong-coupling A-terms and gaugino masses,
and
\bel{vecbnB}
\dot{{\nu}}_r - (\Gammaprime)_{rs}{\nu}_s =
 (\omega^*_u\cdot\dbyd{}{\ln\sigma^*_u})
(\omega_v\cdot\dbyd{}{\ln\sigma_v})
{\gamma}_r
\ee
for $\nu_r$, the associated combinations of soft scalar masses-squared.
In the final approach to the CFT, then, only the least irrelevant
coupling and its associated supersymmetry-breaking couplings will be
nonzero.  The holomorphic soft terms are driven to zero with the same
power law as the coupling is driven to its fixed value.  The soft
masses-squared have a ``driving term'' from the holomorphic terms, but
this shrinks quickly to zero, leading them to be driven to zero also
with the same power law.

However, now consider the effect of the weak and nearly marginal
standard-model couplings $\sigma_{\SM}$.  (We will use subscript $m,n$
to indicate couplings in the strong sector and $\SM$ to indicate weak
marginal ones from the standard model sector.)  Since they are rather
small and slowly running, we may treat them perturbatively.  In
particular, standard model Yukawa and gauge couplings
$y_{\SM},g_{\SM}$ will appear in the
anomalous dimensions $\gamma_m$.  On general grounds, 
\bel{dgamdweak}
\dbyd{\gamma_r}{\ln y_{\SM}} \alt {\rm order}\left({|y_{\SM}|^2\over
16\pi^2}\right) \ ; \
\dbyd{\gamma_r}{(1/\alpha_{\SM})} \alt 
{\rm order}\left({\alpha_{\SM}^2\over 4\pi}\right) \ .
\ee
Thus the contribution of the weak couplings to the flow of the strong
couplings is indeed small.  One might be concerned that
third-generation Yukawa couplings to the Higgs boson might be too
large for this conclusion.  However, any fields which have large
Yukawa couplings at low energy must, in the conformal regime, have
small anomalous dimensions.  Such fields typically couple to the
conformal sector through standard model gauge couplings and irrelevant
operators which run to small values inside the conformal regime.  For
the top quark, for example, it must be that
\bel{dgamdyt}
\dbyd{\gamma_r}{\ln y_t} \alt 
{\rm order}\left({\alpha_{\SM}^2|y_t|^2\over 32\pi^3}\right)
\ee
since by the end of the conformal regime the top quark must couple
to the conformal sector only at two standard model loops.  In short,
all entries in $\Gammaprime$ involving derivatives with respect to
standard model couplings are small.\footnote{This is not absolutely
guaranteed.  If, say, the right-handed top quark superfield is coupled
by a coupling $h$ to an operator $\OO$ in the conformal sector with
dimension very close to two, then $h$ will run slowly and
\eref{dgamdyt} need not be true.  We will assume for simplicity that
$\dim \OO$ is large enough that $h$ becomes small somewhere inside the
conformal regime.}

These derivatives come into the strong-coupling flow equations
\eref{vecblfpB}-\eref{vecbnB} as nearly-constant and small ``driving
terms'' on the right-hand sides.
\bel{vecblfpC}
 \dot{{\ln \sigma_r}} - 
(\Gammaprime)_{rs}{\ln (\sigma_s/\sigma_{s*})}
=  (\Gammaprime)_{r,\SM}{\ln \sigma_{\SM}}+
{\rm order}[(\sigma_s-\sigma_{s*})^2] 
\ee
for the couplings themselves,
\bel{vecbkC}
 \dot{{\omega_r}} -(\Gammaprime)_{rs}{\omega_s} 
= (\Gammaprime)_{r,{\SM}} \omega_{\SM}
\ee
for the associated strong-coupling A-terms and gaugino masses, and
\bel{vecbnC}
\dot{{\nu_r}} - (\Gammaprime)_{rs}{\nu_s} =
 (\Gammaprime)_{r,\SM}{\nu_{\SM}}+
  (\omega^*_{\SM}\cdot\dbyd{}{\ln\sigma^*_{\SM}})
(\omega_{\SM}'\cdot\dbyd{}{\ln\sigma_{\SM'}})
{\gamma}_r
+ {\rm order}\left(\omega_s^2,\omega_s\omega_{\SM}\right)
\ee
for the soft scalar masses.  In short, the weak couplings and
associated soft terms serve as weak driving forces --- of order
$\alpha_{\SM}^2/4\pi$ or of order $|y|^2/16\pi^2$ for light fermions ---
on the
damped strong couplings and associated soft terms.  To see how large
these driving forces are, we must also understand what happens to the
soft terms $\omega_{\SM}$ and $\nu_{\SM}$.

The couplings which do not go to fixed values have beta functions
which are not controlled by the matrix $\Gammaprime$.  They are simply
$$
\beta_{\ln \sigma_{\SM}} = \half \gamma_{\SM} \ .
$$
For all standard model couplings, these anomalous dimensions must be
positive in the limit all standard model gauge couplings are zero, by
unitarity; and so, if negative, must be of order $-\alpha_{\SM}$.
However, if positive, they may be of order one, in which case the
associated couplings run quickly toward zero.  This fact was used to
obtain the fermion mass hierarchy in \cite{Nelson:2000sn}.  In such
models, the
light fermions have large anomalous dimensions near the fixed point,
and their associated Yukawa couplings $y_{\SM}$ have large
positive beta functions.

By contrast, the soft terms are still controlled by $\Gammaprime$.
For A-terms and gaugino masses associated to such couplings
(which include the usual MSSM A-terms and gaugino masses) 
\bel{vecbkD}
 \dot{\omega}_{\SM} =(\Gammaprime)_{\SM,\SM'}{\omega_{\SM'}} 
+ (\Gammaprime)_{\SM,r}{\omega_r} 
\ee
The first term on the right-hand side is small because
$(\Gammaprime)_{\SM,\SM'}$ is perturbative in standard model
couplings.  The second term will be small because $\omega_r$ is only
nonzero (in the conformal regime) due to the driving force in
\eref{vecbkC}, which is proportional to weak couplings.  Thus,
$\omega_{\SM}$ runs slowly, which implies that for standard model
gauge and Yukawa couplings, $M/\alpha$ and $A/y$ are nearly
renormalization-group invariant in the conformal regime.    Thus, for
those couplings which run slowly, the A-terms and gaugino couplings
will remain of the same order as their initial values.  However, for
those Yukawa couplings $y_{\SM}$ which in our models run rapidly to
zero due to large anomalous dimensions, the couplings $A_{\SM}$ run
quickly also, at nearly the same rate.  This gives us our first
important result: {\it all A-terms for light-fermion Yukawa couplings
are roughly proportional to the associated Yukawa couplings.}  The
reason for the proportionality is ``rough'' is that our initial
conditions in the ultraviolet have no special structure.  We assume
only that all Yukawa couplings (A-terms) are of the same order in the
far ultraviolet, with no particular relations among them.  The lack of
precise proportionality leads to the flavor-violating signatures
discussed in section~6.

We may now return to \Eref{vecbkC}.  The driving term on the
right-hand side contains $\omega_{\SM}$, which as we have just seen is
always of order $M_{SUSY}$ times weak couplings.  From
\eref{dgamdweak} and \eref{dgamdyt} it is easily seen that 
the driving term in \eref{vecbkC} is at most of order 
$(\alpha_{\SM}/4\pi) M_{\SM}$, where $M_\SM$ is a standard
model gaugino mass.  Since in the strong sector
$(\Gammaprime)_{rs}\sim 1$, each $\omega_r$ is driven to an
energy-independent value
\bel{omegafixed}
\omega_r \sim  (\Gammaprime)_{r\SM}{\omega_\SM} \sim 
{\alpha_{\SM}\over 4\pi} M_{\SM}\ .
\ee
This shows our earlier assumption that $\omega_r$ makes a small
contribution to the running of standard model soft terms $\omega_{\SM}$ is
consistent.

 Similarly, it is now straightforward to see that the
soft scalar masses-squared associated to the strong couplings
will therefore not run to zero, but will instead reach approximate
fixed points due to the small driving force from the
second term in \Eref{vecbnC}
\bel{nufixed} \nu_r \sim {\alpha_{\SM}\over 4\pi} M_\SM^2 \  
\ee 
where again $M_\SM$ is a standard model gaugino mass.\footnote{Note that it is
always $\alpha_3$ and $M_3$ which appear here, even when the sfermion
whose mass is in question is color-neutral; this is because the
anomalous dimensions of standard model fields $X$
contain strongly-coupled loops with colored
fields $Q$ in them, so derivatives of the anomalous dimensions with
respect to $\alpha_3$ are of order $\alpha_3$ with no additional
suppression.  However, there is not much difference between the
standard model gauge couplings at the scale $M_<$ where this is to be
evaluated.}  Thus, the effect of the nonzero standard model couplings
is to prevent the strong-coupling fixed point from driving the
strong-coupling A-terms, gaugino masses, and soft scalar masses
strictly to zero.  For our purposes, the most important result is the
last one: 
{\it the soft scalar masses for light fermions are not
driven strictly to zero, but instead are each driven to values of
order $\alpha_{\SM}/4\pi$ times the square of the gaugino
mass. Furthermore, these remaining small terms have no flavor symmetry
relating them to one another.}  
This means that flavor symmetry is not
exactly restored at the fixed point, even in models where for
$\alpha_\SM= y_\SM=0$ the CFP would have driven all standard model
scalar masses to zero.  This will also lead to potentially observable
flavor-violating effects, as we discuss in section 6.

As argued in \cite{Nelson:2000sn},  at a CFP we can use the accidental
infrared R invariance to define a
basis for the matter superfields of the standard model.
We now discuss the fixed point running of the  terms in the scalar
mass matrices which are off diagonal in the basis of fields with
definite fixed point $R$ charge. 
We will argue that the result is very simple: these run rapidly
to zero and, when the other
couplings are at their supersymmetric fixed point values, the off-diagonal
mass-squared terms run {\it only} via wave
function renormalization. 

The general result for the running of the 
non-holomorphic mass squared terms is
\bel{tim}
\beta_{{\tim}{}^i_j}= (\bar D_1 D_1+ D_2)\gamma^i_j\ .
\ee
At first
one might be concerned that  at the fixed point an offdiagonal mass term
does not run at
all, since its beta function is proportional to
 derivatives with respect to coupling constants    of off-diagonal
anomalous
dimensions, and the latter vanish at the fixed
point. 
However a
subtlety of    \Eref{tim} is that the operators $D_1$ and
$D_2$ 
contain all {\it possible} couplings, whether or not these
vanish. Usually it is sufficient to consider only the nonzero
couplings, however in some cases   turning
on a coupling $\delta $  could 
change the anomalous dimension matrix by an amount which is linear in
$\delta$,
and so partial derivatives of 
$\gamma^i_j$ with respect to $\delta $ do not vanish.
In particular the operator $D_2$ contains a contribution from
superpotential couplings
$$D_2={\rm gauge\ terms}+{1\over 2}(\tim)^i_j\left(
\yia{\partial\over \partial \yja}+ {\rm perms}+\yjas
{\partial\over \partial \yias }+ {\rm perms}
\right)\ .$$
Here $\yia$ is the superpotential coefficient of $\phi_i\phi_{a_1}
...\phi_{a_n}$. 
In evaluating $D_2$, all  superpotential couplings $\yia$
must be considered
which could produce  contributions to $\gamma^i_j$ which are linear in
$\yia$. 
Assuming all  terms other than the $\tim$ have reached their fixed
point
values (which means that
$D_1$ and the gauge contribution to $D_2$  vanish) we  can find the
beta function for $({\tim})^i_j$.  For the case we are interested in 
  $\phi_{i},\phi_j$ are standard model superfields with   superpotential
couplings 
$$y_i\phi_i{\OO_i}+y_j\phi_j{\OO_j}$$ with $y_i,y_j$ 
nonvanishing at the fixed point. The
$\OO_i$ are  operators of the superconformal sector with {\it different} $R$
charges. 
At the fixed point $\phi_i$ 
 acquires a non vanishing anomalous dimension $\gamma^i_i$, and
the anomalous dimension of 
$\phi_j$ is $\gamma^j_j$.
    If we turn on  infinitesimal  couplings
$$\delta_j\phi_j{\OO}_i+\delta_i\phi_i{\OO}_j$$ then we slightly
perturb the fixed point and different linear combinations of fields
will acquire definite anomalous dimensions. For infinitesimal $\delta$'s,
$\phi_i+{\delta_j\over y_i}\phi_j$ and $\phi_j +{\delta_i\over y_j}\phi_i$
will diagonalize the anomalous
dimension matrix and will get anomalous dimensions $ \gamma^i_i$ and
$\gamma^j_j$ respectively. Hence, 
$${\partial \gamma^i_j\over \partial \delta_j}= {1\over
y_i}\gamma^i_i\ ,$$ and
$${\partial \gamma^i_j\over \partial \delta_i}= {1\over
y_j}\gamma^j_j\ .$$ These do not vanish at the fixed point.
Thus at the fixed point
$$\beta_{\tim{}^i_j}={1\over 2}(\tim)^i_j
\left(y_i {\partial \gamma^i_j\over \partial \delta_j} +
y_j {\partial \gamma^i_j\over \partial \delta_i}+{\rm h.c.}\right)=
(\tilde m^2)^i_j (\gamma^i_i+\gamma^j_j )\ ,$$
and  the off-diagonal scalar mass terms 
run via wave-function renormalization only. 

To summarize, we expect the following spectrum to emerge from the
conformal regime, where the only substantial violation of conformal
symmetry is due to third-generation Yukawa couplings and standard
model gauge couplings.  The standard model gaugino masses $M_\SM$ are
somewhere below $M_{SUSY}$, probably of order $\alpha_{\SM}M_{SUSY}$.
The standard model A-terms are roughly proportional to standard model
Yukawa couplings times an overall scale $A_0$; for the light fermions, both the
Yukawa coupling and the A-term are driven small at about the same
rate.  As seen in section 6, the overall scale $A_0$ of the A terms must be
relatively low, which fortunately is not inherently unnatural.  The
conformal-sector A-terms and gaugino masses are suppressed far below
$M_{SUSY}$ by small standard model couplings and loop factors.  The
strong-sector combinations of soft-scalar masses-squared, {\it which
by assumption include all of the standard model squark and slepton
masses except a few members of the third generation}, are driven to
small values of order $(\alpha_{\SM}/4\pi)M_{\SM}^2$.  Thus, flavor
violation among the partners of the light fermions has been naturally
suppressed.

\section{Consequences of escaping}
At some point a certain relevant coupling in  the model
must drive the theory away from its fixed point.  Since it
is relevant, it is dimensionful at the fixed point.  For
illustration, let us take it to be an effective mass  $m$ for
one or more particles.  The dimension{\it less} coupling
which characterizes its effect is $m/\mu$, which is negligible
at high scales.  Once we approach the scale $m$, the coupling
becomes important.  As we pass through the mass threshold,
there could be important renormalization effects.  What
must happen, or not happen, to ensure that flavor violation
is not reintroduced?

Clearly, if more than one relevant operator is necessary to get rid of
the fixed point and the couplings of the strong sector to the standard
model, then flavor violation may be reintroduced.  Any region of
energy which is not conformal and in which strong flavor-violating
couplings are still present in the superpotential will ruin the
scenario.  We therefore must require that the departure from the fixed
point and removal of all couplings of the strong sector to the
standard model be both rapid and complete.  This is a model-building
issue which we will not address further here; the requirement is not
difficult to satisfy, as shown by explicit example in
\cite{Nelson:2000sn}, so we do not consider it a major issue.  We will
instead argue that if these conditions are satisfied, the threshold
effects are small enough that the previously discussed properties of
our scenario are preserved.

The argument is simple.  The strongly-coupled sector of the theory is
nearly supersymmetric, since the conformal regime has driven the
supersymmetry-breaking in the strong sector down to a scale
$M_{strong}$ which is much smaller than the original $M_{SUSY}$.  The
escape sector is supersymmetric, and thus cannot increase
$M_{strong}$ unless there is an opportunity for a large
renormalization group effect.  If the escape sector does its work over
a small range of energies then there is simply is no opportunity for a
significant enhancement of supersymmetry breaking.  At worst, flavor
violation can be affected only at order one.  

\end{document}